\documentclass[pre,,showpacs,showkeywords,twocolumn, amsmath, amssymb, nodata,nofootinbib]{revtex4}

\usepackage{graphics}
\usepackage{epsfig}
\usepackage{graphicx}
\usepackage{dcolumn}
\usepackage{bm}
\usepackage[english]{babel}

\begin{document}

\title
      {
       Kinetics of the Crystalline Nuclei Growth in Glassy Systems
       }

 \author{
        \firstname{Anatolii~V.}~\surname{Mokshin}}
        \email{anatolii.mokshin@mail.ru}
        \affiliation{Department of Computational Physics, Institute of Physics,
        Kazan Federal University, 420008 Kazan, Russia}
        \affiliation{Landau Institute for Theoretical Physics, Russian Academy of Sciences, 142432 Chernogolovka, Russia}

\author{
        \firstname{Bulat~N.}~\surname{Galimzyanov}}
        \affiliation{Department of Computational Physics, Institute of Physics,
        Kazan Federal University, 420008 Kazan, Russia}
        \affiliation{Landau Institute for Theoretical Physics, Russian Academy of Sciences, 142432 Chernogolovka, Russia}

\today

\begin{abstract}
In this work, we study the crystalline nuclei growth in glassy
systems focusing primarily on the early stages of the process, at
which the size of a growing nucleus is still comparable with the
critical size. On the basis of molecular dynamics simulation
results for two crystallizing glassy systems, we evaluate the
growth laws of the crystalline nuclei and the parameters of the
growth kinetics at the temperatures corresponding to deep
supercoolings; herein, the statistical treatment of the simulation
results is done within the mean-first-passage-time method. It is
found for the considered systems at different temperatures that
the crystal growth laws rescaled onto the waiting times of the
critically-sized nucleus follow the unified dependence, that can
simplify significantly theoretical description of the
post-nucleation growth of crystalline nuclei. The evaluated
size-dependent growth rates are characterized by transition to the
steady-state growth regime, which depends on the temperature and
occurs in the glassy systems when the size of a growing nucleus
becomes two-three times larger than a critical size. It is
suggested to consider the temperature dependencies of the crystal
growth rate characteristics by using the reduced temperature scale
$\widetilde{T}$. Thus, it is revealed that the scaled values of
the crystal growth rate characteristics (namely, the steady-state
growth rate and the attachment rate for the critically-sized
nucleus) as functions of the reduced temperature $\widetilde{T}$
for glassy systems follow the unified power-law dependencies. This
finding is supported by available simulation results; the
correspondence with the experimental data for the crystal growth
rate in glassy systems at the temperatures near the glass
transition is also discussed.
\end{abstract}

\maketitle

\section{Introduction}
A first-order phase transition starts with formation of the nuclei
of a new phase. In particular, the nascent liquid droplets
represent nuclei of the new (liquid) phase in the case of vapor
condensation. Moreover, the bubbles of vapor are such the nuclei
at liquid evaporation, while crystallization is initiating through
formation of crystalline nuclei. According to classical point of
view~\cite{Skripov_1974,Kelton_1983,Debenedetti,Kashchiev_Nucleation,Kalikmanov_2012},
the nucleus of a new phase is capable to demonstrate steady
growth, when it reaches a critical size (see Fig.~\ref{fig:
growth_scheme}). The initial stage of the nucleus growth proceeds
through attachment of the particles of the parent phase to the
nucleus. When concentration of the growing nuclei become high
enough, this growth regime is replaced by the growth through
coalescence of the adjacent growing nuclei. The corresponding
three processes -- the nucleation, the growth by attachment of the
particles and the growth through coalescence of growing nuclei --
represent general basis of any first-order phase
transition~\cite{Kashchiev_Nucleation}, whilst the characteristic
time scales and rates of the processes are determined by the
thermodynamic conditions as well as by inherent features of a
system dependent, mainly, on the type of interaction between the
structural elements (say, particles) which form a system.

\begin{figure}[h]
\centering
\includegraphics[width=0.4\textwidth]{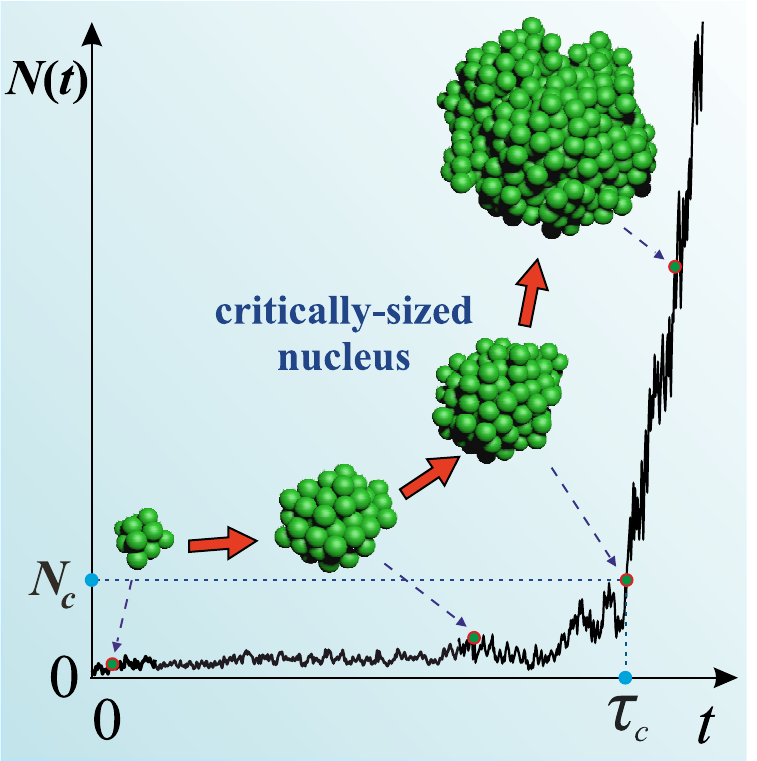}
\caption{Schematic growth trajectory of the largest crystalline
nucleus in a system. Stable growth of a nucleus is possible, when a
nucleus reaches the critical size $N_c$. Here, $\tau_{c}$ is the
waiting nucleation time.}\label{fig: growth_scheme}
\end{figure}

If we restrict our consideration to the crystallization kinetics
of supercooled liquids, then it is possible for the case to
distinguish the inherent features. With increase of the
supercooling level, which can be quantified by $\Delta T/T_m$,
where $T_m$ is the melting temperature and $\Delta T=T_{m}-T$, the
crystal nucleation driving force increases and, thereby, it
accelerates structural ordering in the supercooled liquid. On the
other hand, with the supercooling increase, the structural
transformations is slowing down due to increase of the viscosity.
At the supercooling level deep enough to correspond to the glass
transition temperature $T_g$, the viscosity of a supercooled
liquid takes $\eta \simeq 10^{12}$~Pa$\cdot$s. As a consequence of
high viscosity, the structural transformations in a glass occur so
slow, that \textit{overall crystallization} of the glass at a
temperature $T \ll T_g$ is almost unobservable over acceptable
time scale, albeit the separate crystal nucleation events still
appear~\cite{Fokin/Yritsin_2003}. Moreover, extremely small sizes
and low concentrations of the nascent crystalline nuclei as well
as very low rates of crystal nucleation and of crystal growth
complicate the study of the initial regimes of the crystallization
in glasses by conventional experimental methods. Thus, if the
growth of crystalline clusters in supercooled fluids at low and
moderate levels of metastability is sufficiently well-studied
subject, then there are still many disputed issues related with
the structural ordering in
glasses~\cite{Skripov_1974,Choi_2013,Yang_2015, Rein_Frenkel_1999,Liang_2011,Moore_2011,Bosq_2014,Amstad_2015}. Remarkably, the
techniques based on molecular dynamics simulations appear here to
be fruitful to elucidate microscopic mechanisms of \textit{initial
stages of nucleation and growth processes} for such the
thermodynamic conditions where experimental methods have
difficulties~\cite{Kashchiev_Nucleation,Tang/Harrowell_2013,Zhong/Wang_2014,Wattis_1999,Wilding_2014,Lemarchand_2012,Lemarchand_2013}.

In recent paper~\cite{Mokshin/Galimzyanov_JCP_2015}, it was
reported that the structural ordering in two model glassy systems
proceeds through the nucleation mechanism. For both the systems,
the nucleation times evaluated as functions of temperature follow
the unified scaling law, that was also confirmed by experimental
data for the stoichiometric glasses near the glass transition. In
the present study, we extend results of
Ref.~\cite{Mokshin/Galimzyanov_JCP_2015} to consideration of the
growth kinetics of crystalline nuclei in the glassy systems by
focusing mainly on the initial stage of the crystalline nuclei
growth, where the non-stationary effects associated with nucleus
size fluctuations and with strong size- and time-dependencies of
the nucleus growth rate can be very significant. In the study, we
apply a statistical treatment of simulation results for the growth
kinetics, that is realized, in particular, within the suggested
method of inverted averaging of the independent growth
trajectories.  As a result, the growth law with the
characteristics (the growth rate, the growth lag-time, the growth
exponent) is defined  for each considered thermodynamic state of
the systems, and the corresponding theoretical description becomes
possible. Since the growth trajectories are monitored directly
starting from a nucleation event, then both non-stationary and
steady-state growth regimes are accessible for analysis. In
addition, the temperature dependencies of the evaluated rate
characteristics of crystal growth kinetics as well as their
correspondence to the known experimental data are discussed.

\section{Nuclei growth}

\subsection{General definitions}

Let us start with the growth rate of a nucleus, whose size
overcame the critical value. This quantity  can be expressed in
terms of the number of particles $N$ as
\begin{subequations}\label{eq: vel}
  \begin{equation}\label{eq: vel_N}
    \upsilon_N=\frac{dN}{dt}
  \end{equation}
and through the average radius $R$ as
  \begin{equation}\label{eq: vel_R}
    \upsilon_R=\frac{dR}{dt}.
  \end{equation}
\end{subequations}
Both quantities $\upsilon_N$ and $\upsilon_R$ are positive, when the
cluster has a size larger than the critical size, i.e. $N>N_c$ and
$R>R_c$.  Here, $N_c$ is the critical value of the number of particles, which form the nucleus, and $R_c$ is the critical value of the nucleus radius.
Within the notations of the Becker-D\"{o}ring gain-loss
theory for the nucleation, the growth rate $\upsilon_N$ is defined as a
difference of attachment-rate $g^{+}(N)$ and detachment-rate $g^{-}(N)$ coefficients,
i.e.~\cite{Weinberg_2002}
\begin{equation}\label{eq: v_N_1}
  \upsilon_N = g^{+}(N) - g^{-}(N).
\end{equation}
Taking into account the detailed balance condition
\begin{subequations}\label{eq: balance_eq}
  \begin{equation}\label{eq: balance}
    g^{+}(N-1) P(N-1) = g^{-}(N) P(N),
  \end{equation}
where
  \begin{equation}\label{eq: eq}
    P(N) = P_0 \exp \left ( -\frac{\Delta G(N)}{k_B T} \right )
  \end{equation}
\end{subequations}
is an equilibrium cluster distribution,
one obtains 
\begin{equation}\label{eq: vel_3}
  \upsilon_N = g^+(N) -  g^+(N-1)  \exp\left [ -  \frac{\Delta G(N-1) - \Delta G(N)}{k_B T}   \right ],
\end{equation}
known as growth equation of the Becker-D\"{o}ring gain-loss
theory~\cite{Weinberg_2002}. Here, $\Delta G(N)$ is the free
energy cost required to form a nucleus of the size $N$, and $k_B$
denotes the Boltzmann constant, $T$ is the absolute temperature,
and $P_0$ is a pre-exponential factor.  The continuous version of
this equation has the form~\cite{Zeldovich_1943}
\begin{eqnarray}\label{eq: vel_2}
  \upsilon_N &=& g^+(N)\left \{ 1 - \exp\left [ \frac{1}{k_B T} \frac{d \Delta G(N)}{dN}   \right ]   \right \} \\ &+& \frac{d g^+(N)}{dN} \exp\left [ \frac{1}{k_B T} \frac{d \Delta G(N)}{dN} \right ].\nonumber
\end{eqnarray}

Moreover, according to the classical nucleation theory the free energy $\Delta G(N)$ represents the sum of
the negative bulk contribution $-|\Delta \mu| N$ and the positive surface contribution $\alpha_s N^{2/3}$, i.e.
\begin{equation}\label{eq: free energy}
  \Delta G(N) = - |\Delta \mu| N + \alpha_s N^{2/3},
\end{equation}
where $|\Delta \mu|$ is the difference of the chemical potential per phase unite in the melt and the crystal; and $\alpha_s$
is the temperature-dependent coefficient proportional to the interfacial free energy $\gamma_s$.
For the spherical cluster one has $\alpha_s = (36\pi)^{1/3}\gamma_s/\rho_c^{2/3}$ with
$\rho_c$ being the numerical density of the nascent (crystalline) phase. From Eq.~(\ref{eq: free energy})
one finds
\begin{equation}\label{eq: derivative of energy}
  \frac{d \Delta G(N)}{dN} = |\Delta \mu| \left [  \left( \frac{N_c}{N}  \right )^{1/3} - 1  \right ],
\end{equation}
where
\begin{equation}\label{eq: critical size}
  N_c = \left ( \frac{2}{3} \frac{\alpha_s}{|\Delta \mu|} \right )^{3}
\end{equation}
is the critical cluster size. Inserting Eq.~(\ref{eq: derivative of energy}) into Eq.~(\ref{eq: vel_2}) one obtains
\begin{eqnarray}\label{eq: vel_gen}
  \upsilon_N &=& g^{+}(N)\left \{ 1 - \exp \left [  \frac{|\Delta \mu|}{k_B T} \left (  \left( \frac{N_c}{N} \right)^{1/3}  - 1 \right )  \right ]  \right \} \\ &+& \frac{d g^{+}(N)}{dN} \exp \left [  \frac{|\Delta \mu|}{k_B T} \left (  \left( \frac{N_c}{N} \right)^{1/3}  - 1 \right )  \right ]. \nonumber
\end{eqnarray}
Note that relation~(\ref{eq: vel_gen}) has clear physical meaning.
Namely, it indicates that the growth rate $\upsilon_N$ corresponds
to the kinetic rate $g^{+}(N)$ weighted by thermodynamic factor --
the term in curly brackets, which represents the fraction of added
particles that is not removed due to the detachment process.
Hence, Eqs.~ (\ref{eq: vel_3}), (\ref{eq: vel_2}) and (\ref{eq:
vel_gen}) define the growth rate $\upsilon_N$ of the nucleus,
whose size is larger than the critical size $N_c$, and, that is
important, their application is not restricted by a specific shape
of the nucleus or by a supercooling range. At the same time, it
might be useful to consider some conditions, which lead to known
results and models for the growth rate~\cite{Shneidman_1992} (in
Appendix, see the Zeldovich relation (\ref{eq: Zeldovich rel}) for
the growth rate, the Wilson-Frenkel theory with Eqs. (\ref{eq:
WF_rel}) and (\ref{eq: WF_rel_3}), the Turnbull-Fisher model with
general relation~(\ref{eq: TF_complete}) and the Kelton-Greer
extension with Eqs.~(\ref{eq: Kelton_complete}) and (\ref{eq:
Kelton_near})). In particular, for the case of weak
size-dependence of the attachment rate, growth equation~(\ref{eq:
vel_gen}) takes the form:
\begin{equation}\label{eq: vel_gen1}
  \upsilon_N = g^{+}(N)\left \{ 1 - \exp \left [  \frac{|\Delta \mu|}{k_B T} \left (  \left( \frac{N_c}{N} \right)^{1/3}  - 1 \right )  \right ]  \right \}
\end{equation}
which represents usually a basis for various growth models~\cite{Kashchiev_Nucleation}.

\subsection{Growth of nucleus of \textit{near-critical} sizes}

Taking into account Eq.~(\ref{eq: vel_gen}) and the attachment
rate of generally accepted form [see Eq.~(\ref{eq: kin_gen}) in
Appendix]:
\begin{equation}\label{eq: attach_rate}
  g^{+}(N) = g^{+}(N_c) \left ( \frac{N}{N_c}  \right )^{(3-p)/3}, \hskip 1cm 0 < p \leq 3,
\end{equation}
one obtain directly
\begin{widetext}
\begin{equation} \label{eq: growth_N}
  \upsilon_N = \frac{dN}{dt} = g^{+}_{N_c}\left(\frac{N}{N_{c}}\right)^{\frac{3-p}{3}}\left \{ 1  +\left(\frac{3-p}{3N}-1\right)\exp \left [ \frac{|\Delta \mu|}{k_B T} \left ( \left( \frac{N_c}{N} \right)^{1/3}  - 1 \right )\right]\right \},\,
\hskip 1cm 0 < p \leq 3.
\end{equation}
\end{widetext}

\noindent It is important to stress that Eq.~(\ref{eq: growth_N})
follows directly from definition~(\ref{eq: v_N_1}) for the growth
rate and no additional conditions were applied to values of the
critical size and to metastability level as it is done in the
cases of \textit{ad hoc} growth models. This equation as well as
the continuity equation for the evolving size distribution of the
nuclei and the law of conservation of matter are form the set of
equations, which are necessary to be resolved to reproduce
kinetics of an arbitrary realistic first-order phase
transition~\cite{Kukushkin_JEPT_1998}. Here, $g^{+}(N_c)$ is the
attachment rate for the critically-sized nucleus. As known, the
exponent $p$ takes the integer values for well-defined growth
models including the ballistic and the diffusion-limited models
(the corresponding discussion of the issue can be found in
Ref.~\cite{Shneidman_1992}). If the exponent is $p=3$, then one
has the attachment rate independent of nucleus size. For other
limit case with $p=0$, the attachment rate changes with increase
of nucleus size according to $g^{+}(N) \sim N$. The mixed growth
regimes can also arise with the non-integer values of the
parameter $p$~\cite{Kukushkin_review,Liu}. This is seen from
kinetic growth models~(\ref{eq: ballistic}), (\ref{eq: thetta})
and (\ref{eq: kin_gen}) given in Appendix. In particular, it is
quite reasonable to expect that non-integer values of the
parameter $p$ will be at growth of a nucleus of near-critical
sizes, since there is no clearly defined regime of growth at such
sizes, while stochastic effects in the post-nucleation growth are
significant. Finding general growth law $N(t)$, which will exact
solution of Eq.~(\ref{eq: growth_N}) and will be valid for a whole
size domain and for different growth regimes, is difficult task.
Instead of direct resolving of differential Eq.~(\ref{eq:
growth_N}), we apply other method. Namely, the growth law of
nucleus of near-critical sizes can be formally defined by the
corresponding Taylor series expansion:
\begin{widetext}
\begin{equation} \label{eq: taylor_series}
  N(t) = N_c + \left . \frac{d N(t)}{dt} \right |_{t \approx \tau_c} (t - \tau_c) +  \left . \frac{1}{2} \frac{d^2 N(t)}{dt^2} \right |_{t \approx \tau_c} (t - \tau_c)^2 + \left . \frac{1}{6} \frac{d^3 N(t)}{dt^3} \right |_{t \approx \tau_c} (t - \tau_c)^3 + \mathcal{O}(|t-\tau_c|^4).
\end{equation}
\end{widetext}

\noindent Then, from Eqs.~(\ref{eq: attach_rate}) and (\ref{eq: growth_N}) one finds that expansion~(\ref{eq: taylor_series}) is approximated in the following way:
\begin{subequations} \label{eq: growth series}
  \begin{equation}
    N(t) = N_{c} + \sum_{k=1}^{3}A_{k} (t - \tau_{c})^{k},\label{eq: sub1}
  \end{equation}
  with the coefficients
  \begin{equation}  \label{eq: sub2}
    A_{k} \simeq \frac{3-p}{3N_c \; k!} \left ( g^+_{N_c} \right )^k  \left ( \frac{\beta |\Delta\mu|}{3 N_c} \right )^{k-1},
    \end{equation}
     \begin{equation} \nonumber
     \hskip 1cm 0 < p \leq 3,
  \end{equation}
\end{subequations}
where $g^+_{N_c} \equiv g^+(N_c)$ and $\beta = 1/(k_B T)$. As can
be seen from Eqs.~(\ref{eq: growth series}), nucleus growth
defined by Eqs.~(\ref{eq: growth series}) is represented by the
sum of three contributions, according to which the nucleus size
$N(t)$ evolves as $ \sim t $, $ \sim t^{2}$ and $ \sim t^{3}$,
respectively. The coefficient $A_1 = (3-p)g^+_{N_c}/(3 N_c)$ is
the growth factor, which has dimension of an inverse time and
takes the meaning of an effective growth rate at initial growth
regime, i.e. $A_{1}\equiv\vartheta_{c}$. Further, the coefficient
$A_2$ accounts for the growth acceleration effects, whereas the
term $A_3$ quantifies the change of the growth acceleration with
time for a nucleus of near-critical sizes. With Eq.~(\ref{eq:
growth series}) and known quantities $\tau_c$ and $N_c$, one can
evaluate the attachment rate $g^+_{N_c}$, the reduced chemical
potential difference $\beta|\Delta\mu|$ and the growth exponent
$p$ by fit of Eq.~(\ref{eq: growth series}) to an
``experimentally'' measured growth law.

\section{Simulation details}

We consider the growth of crystalline nuclei in two different glass-formers:
the single-component Dzugutov (Dz) system with the potential~\cite{Dzugutov_PRL_1993}
\begin{eqnarray}\label{eq: Dzugutov}
\frac{U_{Dz}(r^{*})}{\epsilon} &=& C \left(r^{*-m}-D\right)\exp\left(\frac{c}{r^{*}-
a}\right)\Theta(a-r^{*}) \\
&+& D\exp\left(\frac{d}{r^{*}-b}\right)\Theta(b-r^{*}), \nonumber \\
& &r^{*}=r_{ij}/\sigma \nonumber
\end{eqnarray}
and the binary Lennard-Jones (bLJ) system $A_{80}B_{20}$, where the
particles interact via the
potential~\cite{Mokshin/Galimzyanov_JCP_2015}
\begin{eqnarray} \label{eq: bLJ}
\frac{U_{bLJ}(r_{\alpha\beta}^*)}{\epsilon_{\alpha\beta}}&=&4\left[(r_{\alpha\beta}^*)^{-12}-
(r_{\alpha\beta}^*)^{-6}\right], \\
& & r_{\alpha\beta}^{*}=r_{ij}^{_{\alpha\beta}}/\sigma_{\alpha\beta}, \nonumber \\
& & \alpha,\beta \in  \{A,B \}. \nonumber
\end{eqnarray}
Numerical values of the parameters for both the potentials are
presented in Tab.~\ref{tab: potentials}. Here, $r_{ij}$ is the
distance between $i$-th and $j$-th particles, $\Theta(\ldots)$ is
the Heaviside step-function. For the case of the bLJ-system with
potential~(\ref{eq: bLJ}), the labels $A$ and $B$ denote the type of
particles, and the semi-empirical (incomplete) Lorentz-Berthelot
mixing rules are applied (see Tab.~\ref{tab: potentials}). The
characteristics of the potentials, $\sigma$ and $\epsilon$, set the
unit distance and the unit energy, correspondingly, whilst the unit
time is $\tau=\sigma\sqrt{m/\epsilon}$; the particles are of the
same mass, i.e. $m=m_A = m_B =
1$~\cite{Mokshin/Galimzyanov_JCP_2015}.~\footnote[1]{Note that the considered here
binary Lennard-Jones system is characterized by different mixing rules in comparison
with the Kob-Anderson and Wahnstr\"om binary Lennard-Jones systems.
Thereby, it differs from these systems, which form very stable glassy states
and which do not crystallize over simulation time scales. }
\begin{table}[h]
\begin{center}
\small
\caption{Parameters of the Dz-potential~(\ref{eq: Dzugutov}) and
bLJ-potential~(\ref{eq: bLJ}).}
\label{tab: potentials}
\begin{tabular}{ccccccc}
\hline
 & & & Dz & & & \\
 \hline
$C$ & $D$ & $m$ & $a$ & $b$ & $c$ & $d$  \\
$5.82$ & $1.28$ & $16$ & $1.87$ & $1.94$ & $1.1$ & $0.27$ \\
 \hline
\end{tabular} \\
\bigskip
    \begin{tabular}{ccccccc}
     \hline
 & & & bLJ & & \\
 \hline
$\sigma_{\alpha\alpha}$ & $\epsilon_{\alpha\alpha}$ & $\sigma_{\beta\beta}$ & $\epsilon_{\beta\beta}$ & $\sigma_{\alpha\beta}$ & $\epsilon_{\alpha\beta}$ \\
$1.0\sigma$ & $1.0\epsilon$ & $0.8\sigma$ & $0.5\epsilon $ & $0.9\sigma$ & $1.5\epsilon$ \\
 \hline
\end{tabular}
\end{center}
\end{table}

Simulations with the time step $\Delta t = 0.005\;\tau$ were
performed in the $\mathcal{N}PT$-ensemble, where the temperature
$T$ and the pressure $P$ were controlled by the Nos\'{e}-Hoover
thermostat and barostat~\cite{Mokshin/Galimzyanov_JCP_2015}. For a
single simulation run, $\mathcal{N}=6\;912$ particles  were
included into a cubic simulation cell of the volume $V=l_x^3$,
$l_x = l_y = l_z \simeq 20\;\sigma$, and the periodic boundary
conditions are imposed onto all the directions. Glassy samples
were generated by fast isobaric cooling with the rate $dT/dt =
0.001\;\epsilon/(k_B \tau)$ of well equilibrated fluid according
to algorithm presented in details in
Ref.~\cite{Mokshin/Galimzyanov_JCP_2015}. As a result, glassy
samples were prepared at temperatures below the glass-transition
temperature $T_g$ along the isobars with the pressure
$P=14\,\epsilon/\sigma^{3}$ for the Dz-system and with the
pressure $P=17\,\epsilon/\sigma^{3}$ for the bLJ-system. The
Dz-system at the pressure $P=14\,\epsilon/\sigma^{3}$ is
characterized by the the melting temperature
$T_{m}\simeq1.51\,\epsilon/k_{B}$ and the glass transition
temperature $T_{g}\simeq0.65\,\epsilon/k_{B}$ at such the cooling
rate~\cite{Roth_Denton_2000}, while for the bLJ-system at the
pressure $P=17\,\epsilon/\sigma^{3}$ one has the melting
temperature $T_{m}\simeq1.65\,\epsilon/k_ {B}$ and the glass
transition temperature $T_{g}\simeq0.92\,\epsilon/k_{B}$.

Note that to perform the statistical treatment of the simulation results, more than fifty
independent samples were generated for each considered ($P,T$)-state of the systems.
\begin{table}[h]
\begin{center}
\caption{Values of the critical size $N_c$ and the average waiting
time of the critically-sized nucleus $\tau_c$. \label{tab:
size_and_time}}
\begin{tabular}{cccc}
 \hline
       & $T$ $(\epsilon/k_B)$ & $N_c$ & $\tau_c$ $(\tau)$   \\
       \hline
       & $0.05$ & $88 \pm 6$ & $372 \pm 60$ \\
       & $0.1$  & $92 \pm 5$ & $340 \pm 55$ \\
Dz     & $0.15$ & $96 \pm 5$ & $305 \pm 40$ \\
       & $0.3$  & $105\pm 6$ & $250 \pm 40$ \\
       & $0.5$  & $108\pm 5$ & $220 \pm 30$ \\
 \hline
 \hline
       & $0.05$ & $55 \pm 3$ & $820 \pm 80$ \\
       & $0.1$  & $57 \pm 4$ & $800 \pm 75$ \\
bLJ    & $0.2$  & $58 \pm 4$ & $795 \pm 65$ \\
       & $0.3$  & $59 \pm 4$ & $785 \pm 60$ \\
 \hline
\end{tabular}
\end{center}
\end{table}

\section{Nucleation parameters and growth curves from molecular dynamics simulation data \label{sec: methods}}

Classical molecular dynamics simulations allow one to obtain
information about positions of all the particles, which generate
the system. To identify the nuclei of an ordered phase for an
instantaneous configuration of the system we apply the cluster
analysis introduced originally in Ref.~\cite{Wolde_JCP_1996} and
based on computation of the local orientational order parameters,
$q_4(i)$, $q_6(i)$, $q_8(i)$, for each $i$th
particle~\cite{Steinhardt_Nelson_1983}. Details of the algorithm
are given in
Refs.~\cite{Mokshin/Galimzyanov_JCP_2014,Mokshin/Galimzyanov_JCP_2015}.
By means of the cluster analysis we obtain for each $\alpha$th
simulation run the time-dependent growth trajectory
$N_{\alpha}(t)$, which defines the number of particles that
belongs to the largest nucleus in the system at time $t$. Here,
$\alpha$ is the label of simulation run for the $(P,T)$-state.
Four growth trajectories $N(t)$  of the largest nucleus from the
independent molecular dynamics simulations are shown in
Fig.~\ref{fig: MFPT}(a). Moreover, from the set of trajectories
$N_{\alpha}(t)$, where $\alpha = 1,\;2,\;3,\; \ldots,\; 50$, for
the $(P,T)$-state, the curve $\bar{t}(N)$ is defined, which is
known as the mean-first-passage-time curve and which characterizes
the average time of the first appearance of a nucleus with given
size $N$. In accordance with the mean-first-passage-time
method~\cite{Hanggi}, the inflection point of this curve gives the
critical size $N_c$ and the average time $\tau_c$ needed to reach
it, i.e. $\tau_c \equiv \bar{t}(N_c)$. As an illustration,
Fig.~\ref{fig: MFPT}(c) shows a typical mean-first-passage-time
curve derived from the growth trajectories including these given
in Fig.~\ref{fig: MFPT}(a). Further, the inverted
mean-first-passage-time curve $N(\bar{t})$  for sizes $N \geq N_c$
and times $t \geq \tau_c$ will reproduce growth law of a nucleus
in system [see Fig.~\ref{fig: MFPT}(b)]. Then, the curve
$N(\bar{t})$ can be used to extract the values of growth
parameters or to test a theoretical model of nucleus
growth~\cite{Mokshin/Galimzyanov_JPCB_2012}.
\begin{figure}[h]
\centering
\includegraphics[width=0.48\textwidth]{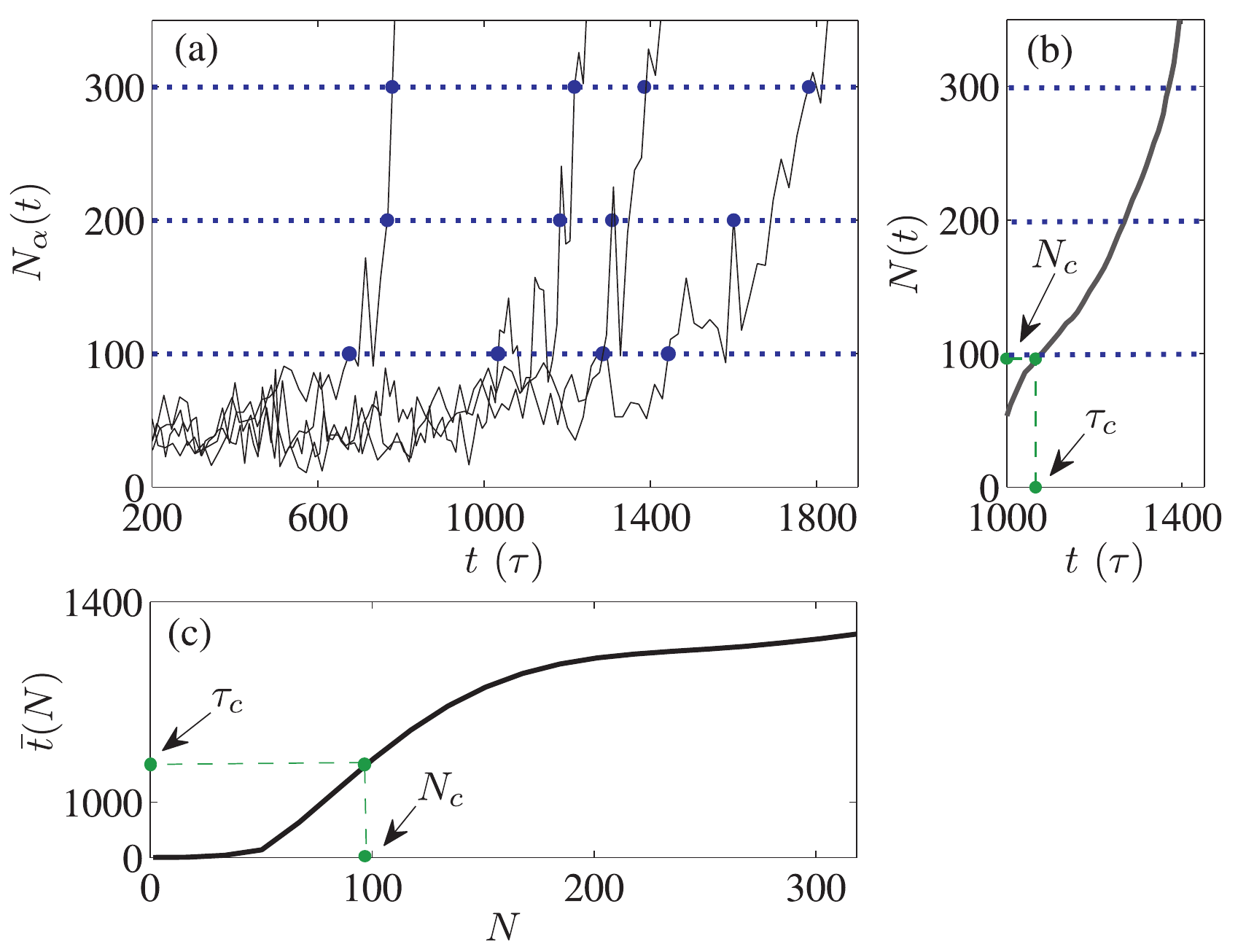}
\caption{ Statistical treatment of the simulation results within
the mean-first-passage-time method. (a) Time-dependent growth
trajectories $N_{\alpha}(t)$ of the largest nucleus extracted from
the data of four independent simulation runs
($\alpha=1,\;2,\;3,\;4$) for the system at a $(P,T)$-state. (b)
Growth law $N(t)$ of the largest nucleus recovered by averaging
over fifty growth trajectories including these four given on panel
(a). (c) Mean-first-passage-time curve defined for the size of the
largest nucleus. The inflection point of this curve defines the
critical size $N_c$ and the average waiting time of the
critically-sized nucleus $\tau_c \equiv \bar{t}(N_c)$.}\label{fig:
MFPT}
\end{figure}

The attachment rate $g_{N_c}^{+}$ can be computed on the basis of
the molecular dynamics simulation data by means of the method
suggested in Ref.~\cite{Wolde_JCP_1996}. Namely, the quantity
$g_{N_c}^{+}$ is defined through the mean-square change of the
nucleus size in the neighborhood of the critical size:
\begin{equation}\label{eq: msd_g}
  g_{N_c}^{+} = \frac{\langle [N(t) - N_c ]^2 \rangle}{2\tau_w},
\end{equation}
where $t \in [\tau_c - \tau_w;\; \tau_c + \tau_w]$ and $\tau_w$ is
the time window, over which the nucleus evolution is
traced~\footnote[3]{In this work, the computation of $g_{N_c}^{+}$
was done on the basis of the data for the largest crystalline
nucleus.}. The angle brackets $\langle \ldots\rangle$ means the
average over independent growth trajectories $N(t)$.

\section{Results}

\subsection{Growth laws}

The cluster analysis reveals that the particles of growing
clusters are located mainly according to fcc structure, while few
amount of the surface particles correspond to hcp structure. This
is observed for both the systems at all the considered
temperatures. Figure~\ref{fig: snap_cluster} demonstrates, as an
example, the crystalline clusters emerging in the Dz- and
bLJ-systems and recognized by means of the cluster analysis.
\begin{figure}[h!]
\centering
\includegraphics[width=0.46\textwidth]{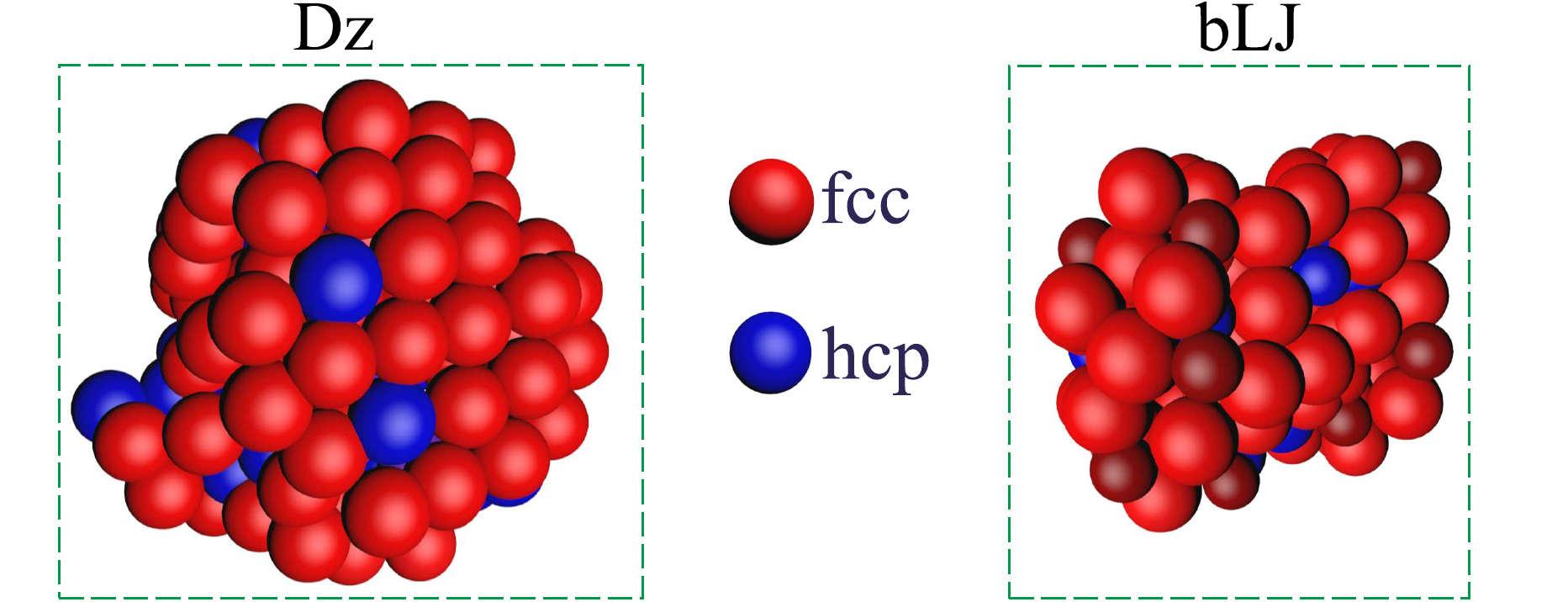}
\caption{Snapshots of the crystalline clusters arising in the
Dz-system at $T=0.5~\epsilon/k_B$ and in bLJ-system at
$T=0.3~\epsilon/k_B$ for which the particles recognized as
belonging to fcc and hcp crystalline phases. } \label{fig:
snap_cluster}
\end{figure}

\begin{figure}[h!]
\centering
\includegraphics[width=0.45\textwidth]{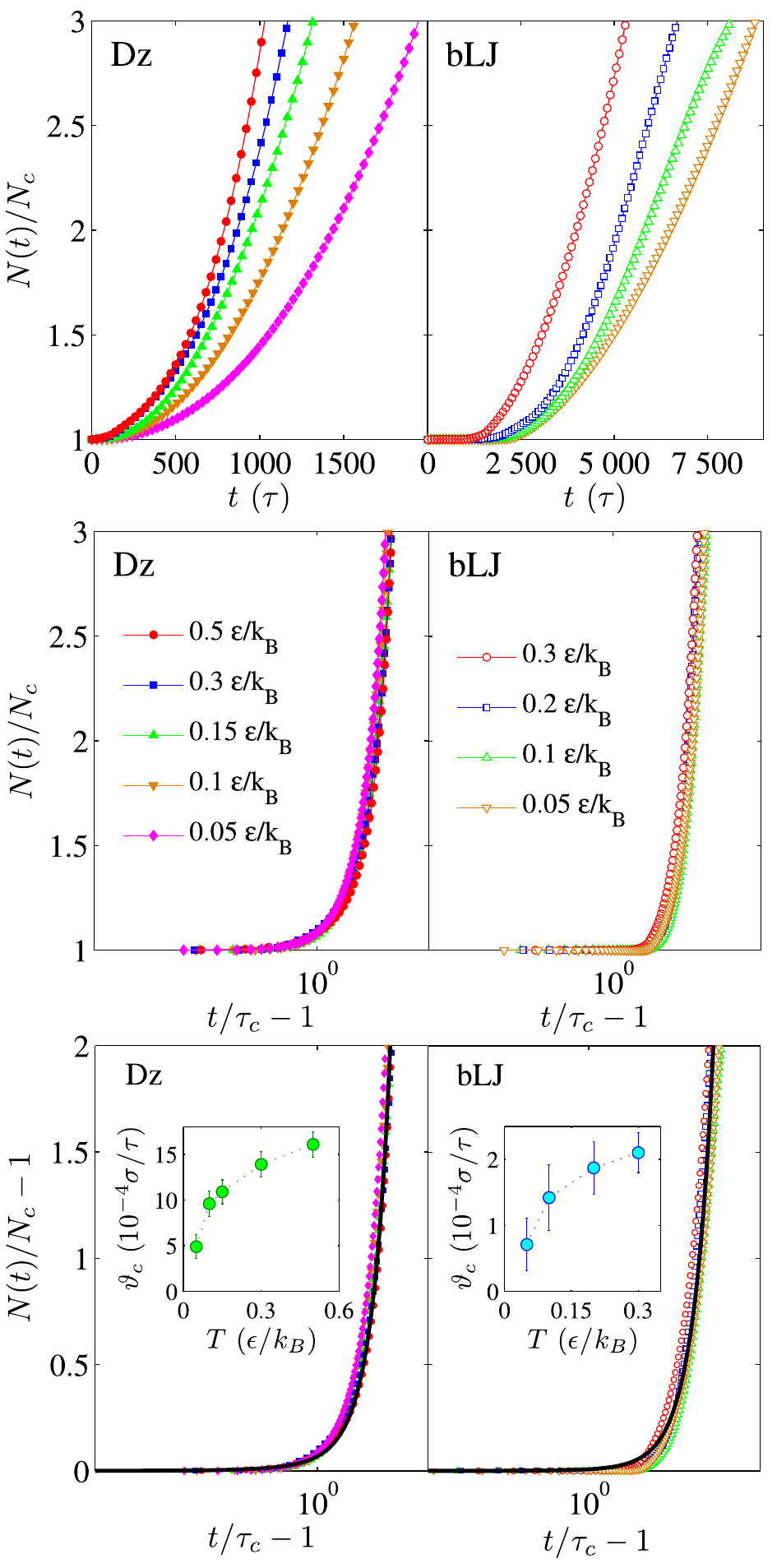}
\caption{\textbf{Top panel}: Growth curves of the crystalline
largest nucleus in glassy Dz- and bLJ-systems at different
temperatures. Each growth curve is a result of the statistical
averaging of the evolution trajectories for the largest nucleus
evaluated from the independent simulation runs. For clarity, the
nucleus size $N(t)$ is rescaled onto the critical size $N_c$.
\textbf{Middle panel}: Growth curves rescaled onto the waiting time
of the critically sized nucleus $\tau_c$. For clarity, the time-axis
is taken in a logarithmic scale. \textbf{Bottom panel}: (Main)
Rescaled growth curves fitted by Eq.~(\ref{eq: growth series}).
Dotted curves correspond to the simulation results; solid curves
represent the fit by Eqs.~(\ref{eq: growth series}). (Insets)
Temperature dependencies of the evaluated growth factor
$\vartheta_c$.}\label{fig: growth_curves}
\end{figure}

The growth curves of the first (largest) crystalline nucleus in the
glassy Dz- and  bLJ-systems at different temperatures are presented
in Fig.~\ref{fig: growth_curves} (top panel). Here, given results
are relevant to the earliest stage of the nucleus growth, where the
nucleus size increases threefold. As it was expected, the growth
process is slowing down with decrease of the temperature $T$.
This is evidenced for both the systems by a shift of the growth
curves with lower temperatures to the domain of longer times.

On middle panel of Fig.~\ref{fig: growth_curves}, the growth
curves are shown in rescaled form, according to which time-scale
is expressed in units of the waiting time $\tau_c$. Moreover, unit
is subtracted from the times in the rescaling in order to relate
the start of the nucleus growth with the zeroth time. As a result
of the rescaling, the growth curves are collapsed onto a
master-curve. The unified behavior of the rescaled growth curves
with respect to the temperature $T$, which is observed
in~Fig.~\ref{fig: growth_curves}, indicates that the crystal
nucleation and growth processes in the glassy systems could be of
the same \textit{kinetic} origin. This means that there are
unified mechanisms of the growth kinetics in the glassy systems,
and the corresponding theoretical description could be done within
a general kinetic model for the growth law. We note that the
similar features were observed before for the growth of the
crystalline nuclei in a model glassy system under homogeneous
shear~\cite{Mokshin/Barrat_2008,Mokshin/Barrat_JCP_2009,Mokshin/Barrat_PRE_2010,Mokshin/Galimzyanov/Barrat_PRE_2013}
as well as for the droplet growth in supersaturated water
vapor~\cite{Mokshin/Galimzyanov_JPCB_2012}.

Taking into account that for each the considered ($P,T$)-state of
both the systems, the attachment rate $g^{+}(N_c)$ is determined
(will be discussed below), while the critical size $N_c$ as well
as the waiting time $\tau_c$ are computed by means of the
mean-first-passage-time
method~\cite{Mokshin/Galimzyanov_JCP_2015}, it becomes possible to
fulfill the fit of the growth curves presented in Fig.~\ref{fig:
growth_curves} (middle panel) by Eqs.~(\ref{eq: growth series})
taking the reduced chemical potential difference $\beta |\Delta
\mu|$ and the growth exponent~$p$ as adjustable
parameters.~\footnote{As shown previously in
Refs.~\cite{Valer_2005,Mosa_2012}, all the particles of a binary
system can be considered without division into subtypes at
relatively small difference in such the characteristics of a
binary system as the partial concentrations, the particle masses
and the particle sizes.} The values of the nucleation
characteristics (the critical size $N_c$ and the average waiting
time of the critically-sized nucleus
 $\tau_c$) determined by means of the method presented in section~\ref{sec: methods} on
 the basis on fifty independent trajectories $N_{\alpha}(t)$ for
 each temperature are given in Table~\ref{tab: size_and_time}.
As can be seen from Fig.~\ref{fig: growth_curves} (bottom panel),
the resulted master-curves of crystal nuclei growth in both the
systems are reproducible by growth law~(\ref{eq: growth series}).
Recall that the presented growth curves are resulted from
averaging over fifty independent growth trajectories. As follows
from our analysis, such the statistics  is not sufficient to
obtain perfect collapse for all the cases over whole considered
time range. Nevertheless, this is quite enough to observe unified
behavior of the growth curves as seen from results of
Fig.~\ref{fig: growth_curves}. Insignificant deviations the
master-curves from growth law~(\ref{eq: growth series}) are
completely covered by small numerical errors of estimated values
of the quantities $\beta |\Delta \mu|$ and $p$. We found that the
growth exponent $p$ takes the value $p = 2.8 \pm 0.15$  for the
Dz-system and $p = 2.99 \pm 0.01$ for the bLJ-system. Note that
the exponent $p$ does not change these values for both the systems
over whole the considered temperature range. The growth law with
the values of the exponent indicates that the attachment rate is
practically independent of the nucleus size, $g^+(N) \simeq
g^+_{N_c}$, at the earliest stage of the nucleus growth. Note that
deviation of values of the parameter $p$ from integers can be due
to the fact that there is no clearly defined regime of nucleus
growth at such sizes, while stochastic effects in the
post-nucleation growth are significant. Further, we find that the
reduced chemical potential  $\beta |\Delta \mu|$ is practically
unchanged for the systems within the considered temperature range.
Namely, one has the values $\beta |\Delta \mu| = 0.215 \pm 0.016$
and $0.167 \pm 0.014$ over the temperature range
$[0.05;\;0.5]$~$\epsilon/k_B$ for the case of the Dz-system, and
$\beta |\Delta \mu| = 0.131 \pm 0.013$ and $0.122 \pm 0.013$
within the temperature range $[0.05;\;0.3]$~$\epsilon/k_B$ for the
case of the bLJ-system.

\subsection{Growth rate}

As seen in Fig.~\ref{fig: growth_curves},  some time after a start
of the nucleus growth, the steady growth regime is established,
where the time-dependent nucleus size  is well interpolated by a
linear dependence. Since time derivative of a growth curve defines
the time-dependent growth rate $\upsilon(t)$ [according to
definition~(\ref{eq: vel_N})], then the derivative  have to
approach a constant value $\upsilon^{(st)}$ for the steady-state
growth regime.~\footnote[2]{Note that the steady-state growth
regime means here a case, in which the growth rate becomes
independent of time.} The growth rate  as function of time,
$\upsilon(t)$, was numerically computed for each the considered
state of the systems on the basis of the growth trajectories
$N(t)$, and the size-dependent growth rate $\upsilon_N$ was
determined from the available data for $N(t)$ and $\upsilon(t)$ by
simple correspondence of the rates $\upsilon$'s and the nucleus
sizes $N$'s at the same time points.

\begin{figure}[h!]
\centering
\includegraphics[width=0.48\textwidth]{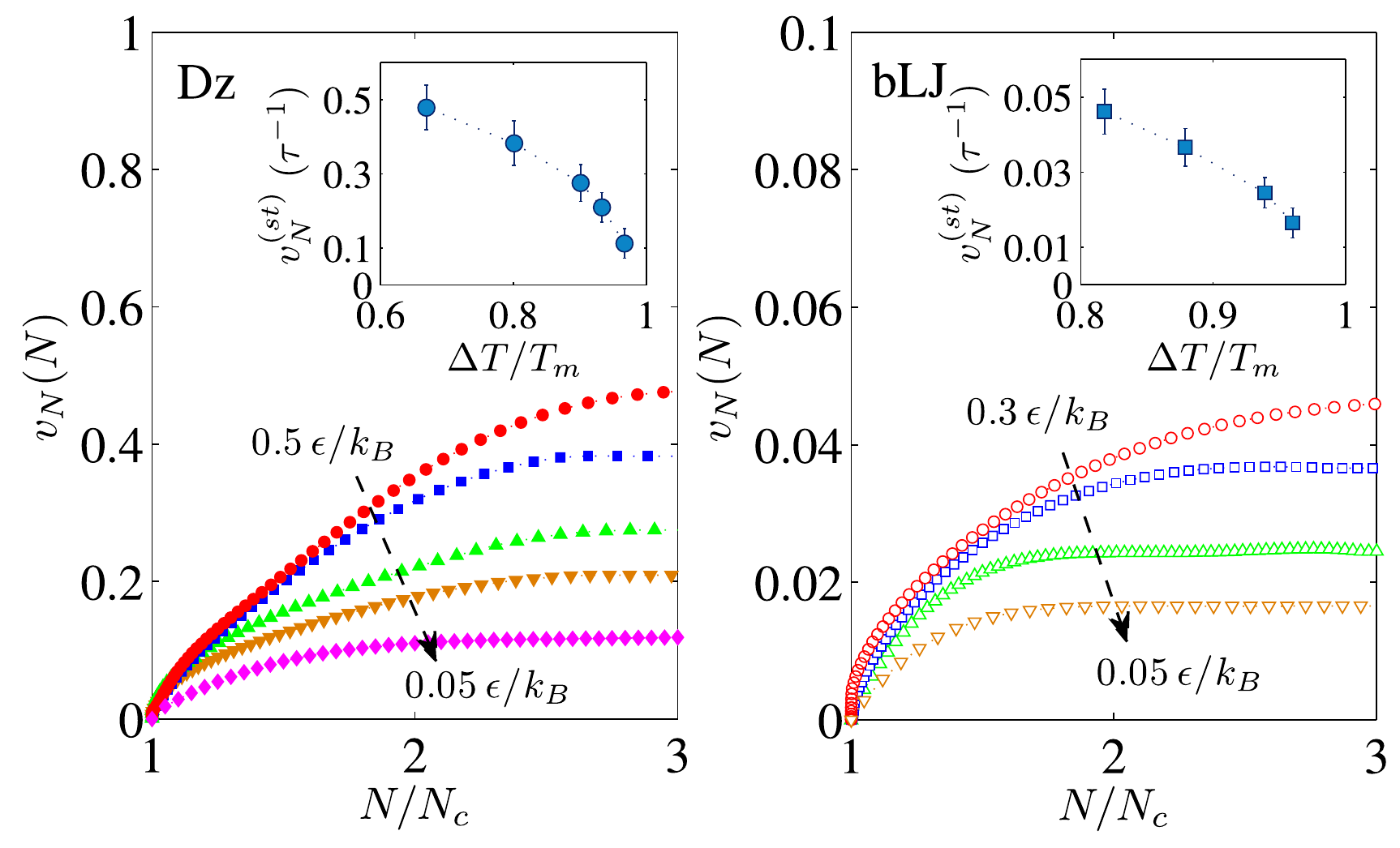}
\caption{\textbf{Main panels:} Growth rate curves $\upsilon_N$
dependent on the crystalline nucleus size for the Dz- and
bLJ-systems at different temperatures. \textbf{Insets:}
Steady-state growth rate $\upsilon_N^{(st)}$ as a function of
$\Delta T/T_m$. Note that the growth law with Eq.~(\ref{eq:
taylor_series}) and its time derivative $dN(t)/dt$ represent, in
fact, the parametric equations for $\upsilon_N(N)$. The values of
the steady-state growth rate $\upsilon_N^{(st)}$ were estimated by
means of numerical solution of the parametric equations.}
\label{fig: growth_vel}
\end{figure}

The computed growth rates as functions of the rescaled size
$N/N_c$ for the systems at different temperatures are presented in
Fig.~\ref{fig: growth_vel}. As expected, the lower growth rates
corresponds to the states with the lower temperatures. One can see
from the figure that the growth rate $\upsilon_N$ increases
initially with the size $N$ and then it reaches the steady-state
value $\upsilon^{(st)}$ for each considered case. Remarkably, the
steady-state growth regime occurs when the cluster size is still
comparable with the critical size $N_c$. For the glassy Dz-system
at the temperatures $T=[0.05;\;0.5]~\epsilon/k_B$ the transition
into a steady-state growth appears at the cluster size $N \simeq
[2.5;\; 3]N_c$, whereas for the glassy bLJ-system at the
temperatures $T = [0.05;\;0.3]~\epsilon/k_B$ one has the
transition at $N \simeq [1.7;\; 3]N_c$. Moreover, the lower
temperature $T$, the smaller value of the cluster size at which
the transition occurs.

The steady-state growth rate
$\upsilon_N^{(st)}$ vs. the supercooling level $\Delta T /T_m$ is
shown in insets of Fig.~\ref{fig: growth_vel}. As seen, the
growth rate $\upsilon_N^{(st)}$ for both the systems decreases with
 increase of the supercooling $\Delta T /T_m$ in such a way that
the growth rates are extrapolated to the zeroth values as we approach
 $\Delta T /T_m=1$ and
$T_0=0$~K. The observed temperature dependence of the
steady-state growth rate $\upsilon_N^{(st)}$
differ from that appears at \textit{low levels of supercooling},
where the growth rate $\upsilon_N^{(st)}$ is proportional to $\Delta
T /T_m$ [see Eqs.~(\ref{eq: equality}), (\ref{eq: WF_rel_3}) and
(\ref{eq: Kelton_near_1}) in Appendix].

As seen from insets of Fig. \ref{fig: growth_vel}, the temperature
dependencies of the steady-state growth rate for both the systems
are similar. If such the behavior is of a common physical origin
for the levels of supercooling, then the behavior has to be
reproducible  within a unified scaling relation, where scaled
values of the  rate characteristic (and of the growth rate) should
be taken as a function of the reduced
temperature~\cite{Mokshin/Galimzyanov_JCP_2015}. For the
temperature range $0 < T < T_m$ corresponding to supercooled
liquid and glass, it looks to be natural to use $T/T_m$ or the
supercooling $(T_m - T)/T_m$ as the reduced temperature. However,
in this case a reasonable consistency in the scaling can be
expected only at the temperatures near
melting~\cite{Fokin/Yritsin_2003}. Moreover, if one uses the
quantity $T/T_g$ as the reduced temperature, as it is doing, for
example, at construction of the scaled structural relaxation time
(or of the scaled viscosity) in the Angell-plot~\cite{Angel_plot},
then the correspondence in time-dependent values  of the rate
characteristic for the different systems will be observed in the
neighborhood of the glass transition temperature $T_g$.
Consequently, the temperatures $\widetilde{T}=T/T_m$ or
$\widetilde{T}=T/T_g$ cannot be considered as convenient
parameters at examination of unified regularities, and a more
adaptive scaling scheme is required.

To compare our results as well as available experimental data we
apply the method presented in
Ref.~\cite{Mokshin/Galimzyanov_JCP_2015}. Main idea of the method
is to display the temperature dependencies of the quantities in
the reduced temperature scale $\widetilde{T}$, where the values of
the zeroth temperature $T_0=0$~K, the glass transition temperature
$T_g$ and the melting temperature $T_m$ are fixed and have same
values $\widetilde{T}_0 = 0$, $\widetilde{T}_g=0.5$ and
$\widetilde{T}_m=1$ for all systems. The correspondence between
the absolute temperatures $T$ and the reduced temperatures
$\widetilde{T}$ is determined
by~\cite{Mokshin/Galimzyanov_JCP_2015}
\begin{subequations}\label{eq: reduced_temp}
  \begin{equation} \label{eq: reduced_1}
    \widetilde{T} = K_1 \left ( \frac{T}{T_g} \right ) + K_2 \left ( \frac{T}{T_g} \right )^2,
  \end{equation}
  with the weight coefficients $K_{1}$ and $K_{2}$:
  \begin{equation}\label{eq: reduced_2}
    K_1 + K_2 = 0.5,
  \end{equation}
  \begin{equation}\label{eq: reduced_3}
    K_1= \left (  \frac{0.5 - \displaystyle{\frac{T_g^2}{T_m^2}}}{1- \displaystyle\frac{T_g}{T_m}} \right ),\ \ \
    K_2 = \left (  \frac{\displaystyle\frac{T_g}{T_m}- 0.5}{\displaystyle\frac{T_m}{T_g}- 1}  \right ).
  \end{equation}
\end{subequations}
Here, the melting temperature $T_m$ and the glass transition temperature $T_g$ are input parameters,
which are taken in the absolute units.

Following Ref.~\cite{Mokshin/Galimzyanov_JCP_2015}, we assume that
the $\widetilde{T}$-dependence of the growth rate
$\upsilon_R^{(st)}(\widetilde{T})$ [and/or
$\upsilon_N^{(st)}(\widetilde{T})$] obeys the power-law
\begin{equation}\label{eq: power_law_vR}
  \upsilon_R^{(st)}(\widetilde{T}) = \upsilon_R^{(g)} \left (  \frac{\widetilde{T}}{\widetilde{T}_g}  \right )^{\chi},
\end{equation}
where  $\widetilde{T}_g = 0.5$ and $\upsilon_R^{(g)}$ is the growth
rate of the crystalline nucleus at the
temperature $T_g$. The exponent $\chi > 0$ characterizes
glass-forming properties of the system. Namely, it takes small
values for the case of the systems, which are not capable
to retain a disordered phase, and, vice versa, the exponent $\chi$
must be characterized by high values for the good
glass-formers~\cite{Mokshin/Galimzyanov_JCP_2015}. In this regard it
is relevant to mention the recent work of Tang and
Harrowell~\cite{Tang/Harrowell_2013}, where the maximum in the crystal growth rate is considered as a quantity correlated with
the glass-forming ability. Since the maximum of $\upsilon_R^{(st)}$
is defined by the variation of the growth rate with the temperature
(with the undercooling), then the exponent $\chi$  in relation
(\ref{eq: power_law_vR}) can be considered as a simple
measure of the glass-forming ability. The validity of
relation~(\ref{eq: power_law_vR}) is easily tested by mapping
the values of the rescaled growth rate into the logarithmic scale as
$(1/\chi)
\log_{10}[\upsilon_R^{(st)}(\widetilde{T})/\upsilon_R^{(g)}]$. In
such representation, the parameter $\chi$ corrects the slope of the
data in $\widetilde{T}$-dependence and the value of $\chi$ is
adjusted so to put the data onto the master-curve
\begin{equation}\label{eq: master-curve}
  \frac{\upsilon_R^{(st)}(\widetilde{T})}{\upsilon_R^{(g)}} = \frac{\widetilde{T}}{\widetilde{T}_g}.
\end{equation}

\begin{figure}[h!]
\centering
\includegraphics[width=0.45\textwidth]{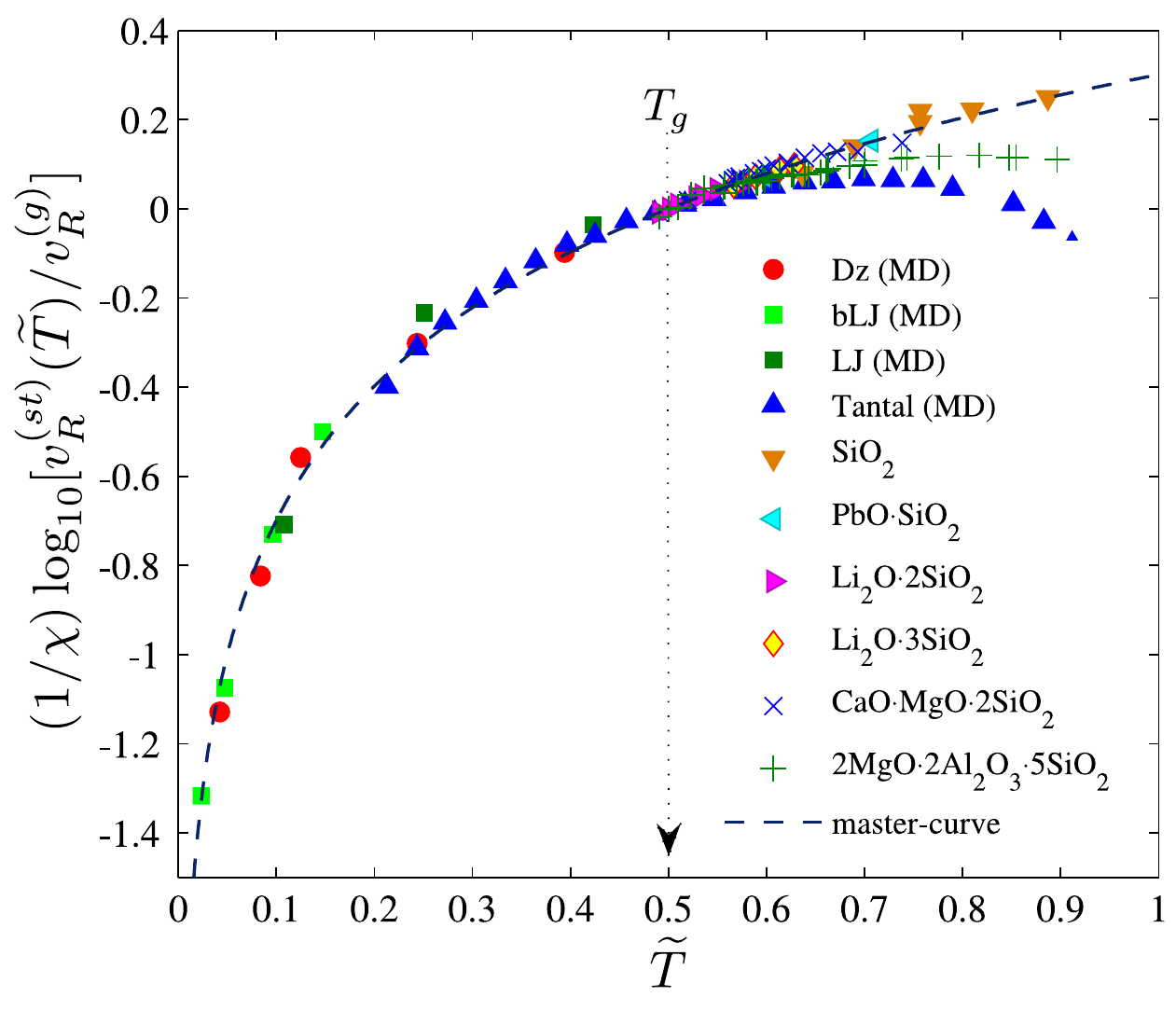}
\caption{Scaled growth rates $\upsilon_R^{(st)}$ vs. reduced
temperature $\widetilde{T}$. Such the plot allows one to compare
the simulation results for the glassy Dz- and bLJ-systems, the
simulation results for the supercooled LJ-system~\cite{Broughton_1982} and tantal~\cite{Zhong/Wang_2014},
the experimental data for the supercooled
SiO$_2$~\cite{Nascimento/Zanotto_2010},
PbO$\cdot$SiO$_2$~\cite{Nascimento/Zanotto_2010},
Li$_{2}$O$\cdot2$SiO$_{2}$~\cite{Fokin/Yritsin_2003},
Li$_{2}$O$\cdot3$SiO$_{2}$~\cite{Ogura/Hayami_1968},
CaO$\cdot$MgO$\cdot2$SiO$_{2}$~\cite{Reinsch/Nascimento_2008} and
$2$MgO$\cdot2$Al$_2$O$_3\cdot
5$SiO$_2$~\cite{Reinsch/Nascimento_2008}. The dashed line
reproduces the master-curve~(\ref{eq: master-curve}). Arrow
indicates the scaled glass transition temperature $\widetilde{T}_g
= 0.5$. For the glassy Dz- and bLJ-systems, values of
$\upsilon_R^{(g)}$ were defined by extrapolation of the data
$\upsilon_R^{(st)}(\widetilde{T})$ to the temperature
$\widetilde{T}_g=0.5$. Values of the parameters $\chi$ and
$\upsilon_R^{(g)}$ are given in Tab.~\ref{Tab: parameters}.}
\label{fig: Scaled_T}
\end{figure}

The temperature dependencies of the growth rates evaluated for case
of the Dz- and bLJ-systems as well as estimated from molecular
dynamics simulations of crystallized single-component LJ-system~\cite{Broughton_1982} and tantal~\cite{Zhong/Wang_2014}
and the available experimental data for the crystal growth rates
$\upsilon_R^{(st)}$ for SiO$_2$~\cite{Nascimento/Zanotto_2010},
PbO$\cdot$SiO$_2$~\cite{Nascimento/Zanotto_2010},
Li$_2$O$\cdot$2SiO$_2$~\cite{Fokin/Yritsin_2003},
Li$_2$O$\cdot$3SiO$_2$~\cite{Ogura/Hayami_1968},
CaO$\cdot$MgO$\cdot$2SiO$_2$~\cite{Reinsch/Nascimento_2008},
2MgO$\cdot$2Al$_2$O$_3\cdot$5SiO$_2$~\cite{Reinsch/Nascimento_2008}
are given in the same Fig.~\ref{fig: Scaled_T}. The defined
values of the parameter $\chi$ as well as the values of
the parameter $\upsilon_R^{(g)}$ for the considered systems are
presented in Tab.~\ref{Tab: parameters}.
\begin{table*}[ht]
\small
\begin{center}
\caption{The melting temperature $T_m$, the glass transition
temperature $T_g$, the steady-state growth rate $\upsilon_R^{(g)}$
at the transition temperature $T_g$,  the exponent $\chi$ found from
Eq.~(\ref{eq: power_law_vR}), the attachment rate $g_{N_c}^{(g)}$ at
the glass transition temperature $T_g$, the exponent $\varsigma$
evaluated from fit of Eq.~(\ref{eq: scaled_g+}) to the simulation
results.}
\label{Tab: parameters}
\begin{tabular}{cccccccc}
\hline
 & System & $T_m$ & $T_g$ & $\upsilon_R^{(g)}$ & $\chi$ & $g_{N_c}^{(g)}$ & $\varsigma$  \\
\hline
& Dz (at $P=14\varepsilon/\sigma^3$) & $1.51\;\varepsilon/k_B$ & $0.65\;\varepsilon/k_B$ & $(23.8 \pm 2.7) \cdot 10^{-4}\sigma/\tau$ & $0.37 \pm 0.06$ & $(13.9 \pm 1.6)\cdot\tau^{-1}$ & $0.31 \pm 0.04$ \\
& bLJ (at $P=17\varepsilon/\sigma^3$)& $1.65\;\varepsilon/k_B$ & $0.92\;\varepsilon/k_B$ & $ (4.7 \pm 0.7) \cdot 10^{-4}\sigma/\tau$ & $0.41 \pm 0.07$ & $(13.1 \pm 1.5)\cdot\tau^{-1}$ & $0.58 \pm 0.06$  \\
\cite{Broughton_1982} & LJ & $0.62\;\varepsilon/k_B$ & $0.4\;\varepsilon/k_B$ & $0.41 \pm 0.05\;\sigma/\tau$ & $0.39\pm0.07$ & -- & -- \\
\cite{Zhong/Wang_2014} & Ta                                 & $3\;290$\;K & $1\;650$\;K & $41.5\pm4$\;m/s    & $3.5 \pm 0.6$  & -- & -- \\
\cite{Nascimento/Zanotto_2010} & SiO$_2$                            & $2\;000$\;K & $1\;450$\;K & $(1.1\pm0.2)\cdot 10^{-12}$\;m/s & $12.6\pm 1.4$  & -- & -- \\
\cite{Nascimento/Zanotto_2010}  & PbO$\cdot$SiO$_2$                 & $1\;037$\;K & $673$\;K  & $(1.5\pm0.2)\cdot 10^{-11}$\;m/s & $22.6\pm 1.5$  & -- & -- \\
\cite{Fokin/Yritsin_2003} & Li$_{2}$O$\cdot2$SiO$_{2}$        & $1\;306$\;K & $727$\;K & $(2.2\pm0.4)\cdot 10^{-12}$\;m/s & $49.2\pm 4.5$  & -- & --   \\
\cite{Ogura/Hayami_1968} & Li$_{2}$O$\cdot3$SiO$_{2}$         & $1\;306$\;K & $734$\;K & $(1.1\pm0.3)\cdot 10^{-11}$\;m/s & $34.5\pm 2.5$  & -- & --   \\
\cite{Reinsch/Nascimento_2008} & CaO$\cdot$MgO$\cdot2$SiO$_{2}$    & $1\;664$\;K & $993$\;K  & $(9.4\pm1.2)\cdot 10^{-14}$\;m/s & $49.8\pm 4.8$  & -- & --   \\
\cite{Reinsch/Nascimento_2008} & $2$MgO$\cdot2$Al$_2$O$_3\cdot 5$SiO$_2$ & $1\;740$\;K & $1\;088$\;K & $(3.9\pm0.8)\cdot 10^{-12}$\;m/s & $53.2\pm 6.2$  & -- & --   \\
\hline
\end{tabular}
\end{center}
\end{table*}

As seen from Fig.~\ref{fig: Scaled_T}, the growth rates for all
the systems follow the master-curve~(\ref{eq: master-curve}) for
the temperatures $\widetilde{T}$ near and below the glass
transition temperature $\widetilde{T}_g=0.5$. Numerical values of
the parameters $\chi$ and $\upsilon_R^{(g)}$ are given in
Tab.~\ref{Tab: parameters}. This directly indicates that behavior
of the growth rate $\upsilon_R^{(st)}(\widetilde{T})$ for the
temperature range, $\widetilde{T} \leq 0.5$, is reproducible by
the power-law and relation~(\ref{eq: power_law_vR}) is unified for
the $\widetilde{T}$-dependent growth rates of the systems. This
result is consistent with the findings of
Ref.~\cite{Mokshin/Galimzyanov_JCP_2015}, where it was
demonstrated that the scaled crystal nucleation time $\tau_1$ in
glassy systems as a function of the reduced temperature
$\widetilde{T}$ follows the unified power-law dependence $\tau_1
\sim (1/\widetilde{T})^{\gamma}$. In addition, the power-law
temperature dependence of the structural relaxation time
$\tau_{\alpha} \sim (1/\widetilde{T})^{\gamma}$ generalizes the
Avramov-Milchev model for the
viscosity~\cite{Mokshin/Galimzyanov_JCP_2015}, while the exponent
$\gamma$ is related with the fragility index $m$. Therefore, it
quite reasonable to anticipate that such the temperature
dependence is generic for the rate characteristics of structural
transformations in glasses, where the inherent kinetics is
dominant over thermodynamic aspects. This is differ from how
nucleation and growth proceed at the low levels of supercooling,
where impact of the thermodynamic contributions on the values of
the rates is significant and where the growth rate
$\upsilon_R^{(st)}$ increases with increase of
supercooling~\cite{Debenedetti}. Thus, the character of the
$\widetilde{T}$-dependent growth rates for the low supercoolings
diverges from the $\widetilde{T}$-dependence specified by
Eq.~(\ref{eq: power_law_vR}). As can be seen in Fig.~\ref{fig:
Scaled_T}, the behavior of the curve
$\upsilon_R^{(st)}(\widetilde{T})$ for the systems starts to be
different as the temperature $\widetilde{T}$ approaches the
melting point $\widetilde{T}_m=1$, where the growth rate
$\upsilon_R^{(st)}(\widetilde{T}_m)$ is equal to zero for any
system.

\subsection{Attachment rate}

Since the growth kinetics depends directly on the attachment rate,
$g_{N}^{+} \equiv g^+(N)$ [see Eq.~(\ref{eq: vel_gen})],
therefore, it is important to consider how the quantity
$g_{N}^{+}$ behaves with the temperature variation, that can be
done with the term $g_{N_c}^{+}$, which is the attachment rate for
the critically-sized nucleus. We recall here that some theoretical
models of the growth kinetics (e.g., the Turnbull-Fisher model
with relation~(\ref{eq: TF_complete}) as well as the Kelton-Greer
extension with relation~(\ref{eq: TF_near}), see Appendix) utilize
the quantity $g_{N_c}^{+}$ instead of the size dependent
$g_{N}^{+}$.

\begin{figure}[h!]
\centering
\includegraphics[width=0.38\textwidth]{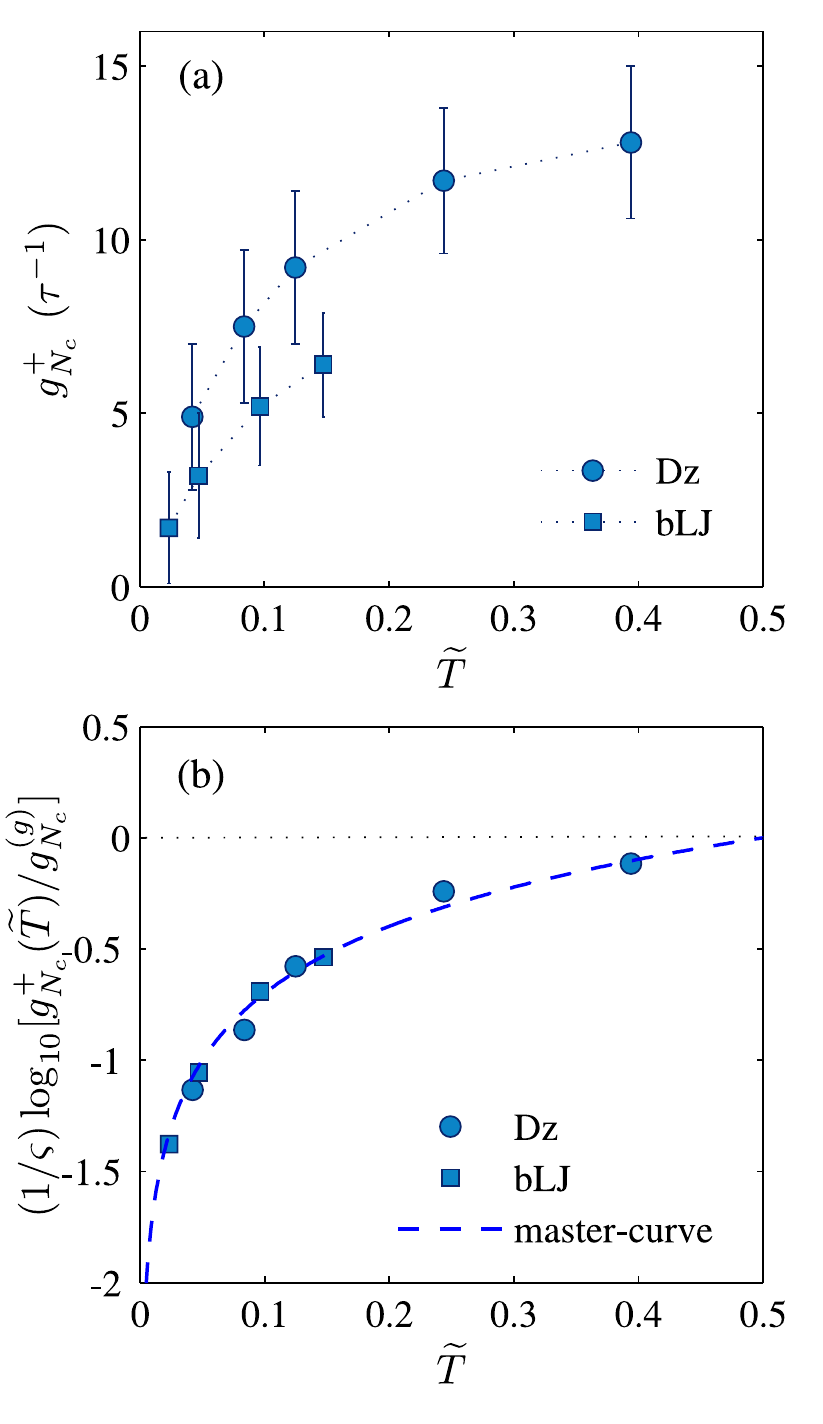}
\caption{(a) Attachment rate for the critically-sized nucleus
$g^{+}_{N_c}$ a a function of the reduced temperature
$\widetilde{T}$ for the Dz-system and for the bLJ-system. (b)
Scaled attachment rate $(1/\varsigma)
\log_{10}[g^{+}_{N_c}(\widetilde{T})/g^{(g)}_{N_c}]$ versus the
reduced temperature $\widetilde{T}$. The values of the attachment
rate $g^{(g)}_{N_c}$ for the systems are evaluated by
extrapolation data $g^{+}_{N_c}(\widetilde{T})$ to the glass
transition temperature $\widetilde{T}_g$. The dashed line
correspond to the master-curve $g^{+}_{N_c}/g_{N_c}^{(g)}  =
\widetilde{T}/\widetilde{T}_g$. Values of the exponent $\varsigma$
and the quantity $g^{(g)}_{N_c}$ are given in Tab.~\ref{Tab:
parameters}.}\label{fig: Attachment_rate}
\end{figure}

The values of the quantity $g_{N_c}^{+}$ for both the systems are
given in Fig.~\ref{fig: Attachment_rate}(a), where the temperature
is plotted into the reduced scale, $\widetilde{T}$. As can be
seen, the quantity $g_{N_c}^{+}$ decreases with the decrease of
the temperature. Nevertheless, the attachment rate as well as the
growth rate take the finite values for the systems even at the
deep levels of supercooling, and are still detectable over a
simulation time scale. Hence, even insignificant displacements of
the particles may result in structural transformations in
high-density glassy systems, where the particles interact through
an isotropic
potential~\cite{Khusnutdinoff/Mokshin_JNCS_2011,Khusnutdinoff/Mokshin_PhysA_2012,Mokshin/Galimzyanov_JCP_2015}.

By analogy with results for the steady-state growth rate (see Fig.~\ref{fig: Scaled_T}),
the scaled attachment rate versus
the reduced temperature $\widetilde{T}$ is presented in Fig.~\ref{fig: Attachment_rate}(b).
Here, the parameter $g_{N_c}^{(g)}$ is the attachment rate at the glass transition temperature
$\widetilde{T}_g=0.5$, and its numerical values for the systems are estimated by extrapolation
of the attachment rate $g^+_{N_c}(\widetilde{T})$ to the temperature domain near $\widetilde{T}_g$.
The exponent $\varsigma$ characterizes how the attachment rate changes with the temperature, that is similar to physical meaning of the exponent $\chi$.
At the construction of Fig.~\ref{fig: Attachment_rate}(b), the exponent $\varsigma$
is taken as adjustable parameter, which corrects the slope of a curve; while the quantity $g_{N_c}^{(g)}$
is defined in a way to guarantee the zeroth value of $(1/\varsigma)\log_{10}[g_{N_c}^{+}(\widetilde{T})/g_{N_c}^{(g)}]$
at the glass transition temperature $\widetilde{T}_g=0.5$.
The estimated values of the quantities
$g_{N_c}^{(g)}$ and $\varsigma$ are given in Tab.~\ref{Tab: parameters}.
It is seen from Fig.~\ref{fig: Attachment_rate}(b) that the $\widetilde{T}$-dependencies of the obtained
attachment rate $g^{+}_{N_c}$ for both the glassy systems are well reproduced by the power-law
\begin{equation} \label{eq: scaled_g+}
  g^{+}_{N_c}(\widetilde{T}) = g_{N_c}^{(g)} \left ( \frac{\widetilde{T}}{\widetilde{T}_g}  \right  )^{\varsigma}.
\end{equation}
[We recall that the glass transition temperature is $\widetilde{T}_g =0.5$].

Remarkably, the exponent $\varsigma$ takes values close to values
of the exponent $\chi$ characterizing the growth kinetics. This is
not surprising, since  the growth  and attachment rates,
$\upsilon_N$  and $g_{N_c}^{+}$, are expected to be significantly
correlated at the deep levels of supercooling. Taking into account
the defined values of $g_{N_c}^{(g)}$ we deduce that the
attachment rate decreases by two times over the temperature range
from $T_g=0.65\;\epsilon/k_B$ to $T=0.05\;\epsilon/k_B$ for the
Dz-system and by $4.4$ times over the temperature range  from
$T_g=0.92\;\epsilon/k_B$ to $T=0.05\;\epsilon/k_B$ for the
bLJ-system.

\section{Discussion}

Traditional experiments on crystal growth are capable to probe
crystalline nuclei within micron size range~\cite{Weinberg_2002}, and the theoretical
models for the growth kinetics are required to extrapolate the experimental data into the domain of
smaller sizes, where the crystal growth initiated, and to recover overall picture of the nucleation-growth
process~(see, for example, Refs.~\cite{Granasy_JCP_2000,Shneidman_1992}).
In the given study, we apply an opposite handling. On the basis of simulation data for model glassy systems
at different temperatures, the crystal growth is directly restored starting from a nucleation event.
This provides a possibility to define the time-dependent crystal growth law as well as to evaluate a crystal growth rate characteristics as size- and
temperature-dependent terms.

Statistical treatment of the characteristics of growth kinetics is
realized in this study as follows. The growth law is
defined by means of the mean-first-passage-time
method~\cite{Mokshin/Galimzyanov_JCP_2014,Mokshin/Galimzyanov_JPCB_2012},
in which an waiting time scale sets in accordance to the
certain size $N$. As a result, one can
estimate the growth law $N(t)$ with $N \geq N_c$ on
the basis of the independent growth trajectories for the case, when
the direct averaging  $N(t) = \sum N_{\alpha}(t)$ does not work,
because the nucleation time $\tau_c$ can take a value from a range larger than the characteristic time scale of crystal growth [see Fig.~2(a)].
 Within the  statistical treatment, the
nucleation and growth rate parameters are definable within a common
method utilizing the time-dependent growth trajectories
resulted form independent experiments or molecular dynamics
simulations.

The analysis reveals that an envelope of the nuclei at the initial
growth stage is reproduced well by a
sphere~\cite{Mokshin/Galimzyanov_JCP_2015}, while the crystalline faces of the nuclei are not well-defined at
the stage~\cite{Kelton_review}. In the case, the average nucleus
radius coincides with the proper nucleus radius; and one can
directly relate the size parameters -- the number of particles
enclosed in a nucleus $N$ and the nucleus radius $R$. This
allows one to carry out reasonably the treatment of the results by
means of the so-called ``mean radius
approaches''~\cite{Mean_radius}.
The results obtained in the study reveal that the size of
crystalline nuclei in the considered glassy systems evolves with
time at post-nucleation stage according to dependence, which is an approximate solution of the growth equation. Such the
character of the growth of nascent crystalline nuclei holds over
temperature range. As a result, the growth law can be represented in
a unified form with the scaled parameters -- the time $t/\tau_c$ and the size $N/N_c$. We stress that such the character of the growth law is consistent with the limit solution of growth
equation~(\ref{eq: growth_N}), as well as with the theoretical
suggestions~(see p.~378 in Ref.~\cite{Kashchiev_Nucleation}) and
with previous simulation results~\cite{Mokshin/Galimzyanov_JCP_2015,Mokshin/Galimzyanov_JPCB_2012,Mokshin/Galimzyanov_JPCS_2012}.

The growth rate at initial growth stage depends strongly on the
nucleus size. Nevertheless, at high levels of supercooling
considered in the study, transition into the steady-state growth
regime arises, when the nucleus size increases in two-three times
in respect to a critical size. Notable, the nucleus size $R$ equal
to $2R_c$ is also mentioned by Langer~(p.~14 in
Ref.~\cite{Langer_RMP_1980}) and is related assumedly with a
maximum of growth rate $\upsilon_R$ arisen in the pre-dendridic
stage of crystal growth. Hence, the crystalline growth in glasses
can be characterized by a single value of the growth rate
$\upsilon^{(st)}$ over an extended size range, that simplifies
considerably a theoretical description of crystallization kinetics
in the systems~\cite{Skripov_1974,Debenedetti}. Further, according
to  Eq.~(\ref{eq: growth_N}), the size-dependent growth rate
$\upsilon_N$ is defined by the three input quantities: the
attachment rate $g^{+}(N)$, the critical size $N_c$ and the
reduced difference of the chemical potential $|\Delta \mu|/k_B T$.
Herewith, the known theoretical models (see Appendix) are derived
from the equation by an approximation for the $N$-dependent
attachment rate $g^{+}(N)$ [as is done, for example, for the pure
kinetic models and for kinetic extension of the Turnbull-Fisher
model, see Appendix] and/or by considering the system at specific
thermodynamic conditions with the given ratios $|\Delta \mu|/k_B
T$ and $N_c/N$ [as in the Wilson-Frenkel
theory~\cite{Wilson_1900,Frenkel_1932}, see Appendix]. In the
present study, consideration of the growth kinetics was directly
done on the basis of a primary equation of growth kinetics, i.e.
Eq.~(\ref{eq: growth_N}).

According to classical view on the nucleation-growth
process~\cite{Kashchiev_Nucleation}, the temperature
dependence of the growth rate is characterized by a maximum, which
appears due to the features in the $T$-dependence of the kinetic
rate coefficient $g^+(N)$ and the thermodynamic factor -- the
expression in curly brackets in Eq.~(\ref{eq: vel_gen}), for example. Hence, with
decrease of the temperature, the first quantity also decreases,
whereas the second term increases. In the given study, we focus on
the crystalline nuclei growth in glasses at deep levels of
supercooling. At such the thermodynamic conditions, the steady-state
growth rate $\upsilon^{(st)}$ is defined mainly by the kinetic term
$g^+(N)$ and, thereby, it increases with the temperature $T$, that
is opposite to well-known behavior of the growth rate observable
experimentally in the systems at low and moderate levels of
supercooling~\cite{Ogura/Hayami_1968,Reinsch/Nascimento_2008,Nascimento/Zanotto_2010}.

General patterns in the temperature dependencies of the growth rate
detectable for various crystallizing systems~\cite{Reinsch/Nascimento_2008,Nascimento/Zanotto_2010}
indicates that a unified theoretical description of the growth
rate as a function of the temperature as well as direct
comparison of the experimental data for the systems by means of
scaling relations are possible. As an attempt to do
such the comparison, one can mention the corresponding part in
Langer's review~\cite{Langer_RMP_1980} (Fig.~15 on p.~19 and its
discussion), where the growth rates measured for ice and
succinonitrile are compared with the Ivantsov relation. In
particular, it follows from Fig.~15 given in
Ref.~\cite{Langer_RMP_1980} that the dimensionless growth rate
as a function of dimensionless undercooling presented in
double-logarithmic plot is well interpolated by a power-law
dependence. In this work, we extend the idea of a unified
description of the nucleation-growth
kinetics~\cite{Angel_plot,Hale_1986,Hale_2010},
that is realized here on the basis of the reduced temperature
$\widetilde{T}$-scale concept~\cite{Mokshin/Galimzyanov_JCP_2015}.
Note that the reduced temperature $\widetilde{T}$-scale differs from
such the dimensionless temperatures applying usually to characterize
the thermodynamic states of supercooled liquids and glasses as the
undercooling $\Delta T/T_m$~(see Ref.~\cite{Frenkel_book}), or the
dimensionless temperatures $T/T_m$ and
$T/T_g$ (see Ref.~\cite{Angell_CA,Angel_plot}). Namely, the
$\widetilde{T}$-scale ranks the temperature range $0\leq T \leq T_m$
uniformly for supercooled systems, and, thereby, this allows one to
compare the temperature dependent characteristics of the systems,
whose the glass-forming abilities differs significantly. By means of
this approach, we found that the growth rates extracted from
simulations and experimental data for the different systems and
plotted as a function of $\widetilde{T}$ in the temperature range
$0<T\leq T_g$ approximate the power-law dependence with the
exponent, which is using as fitting parameter and could be using for
numerical estimate of the glass-forming ability of a
system~\cite{Tang/Harrowell_2013}. Finally, the attachment rate
evaluated on the basis of simulation data and presented as a
function of the reduced temperature $\widetilde{T}$ follows also the
power-law dependence, which is correlated with the scaling law found
before for crystal nucleation times in the
systems~\cite{Mokshin/Galimzyanov_JCP_2015}.

Divergence of the growth rate values and their deviation from the
universal power-law at the temperatures $T_g < T < T_m$ is  not
surprising. At low levels of supercooling, the thermodynamic
properties have main impact on the nucleation-growth processes.
This is reproducible, in particular, within the Wilson-Frenkel
theory (see Eqs.~(\ref{eq: WF_rel_2}) and (\ref{eq: WF_rel_3}) in
Appendix) and the Turnbull-Fisher model (see Eq.~(\ref{eq:
Kelton_near}) in Appendix). Here, the growth rate as well as the
nucleation rate are increasing with supercooling increase till the
system temperature $T$ will not be comparable with the glass
transition temperature (i.e. $T \leq T_g$), where the nucleation
and growth rates take the highest values. With further increase of
supercooling, the slowing down of the system kinetics is resulted
into decrease of all the transition rate characteristics (namely,
the nucleation, growth and attachment rates). Therefore,
expectedly, these rates can be correlated as functions of the
temperature $T$ at the range  $0< T \leq T_g$. This is supported
by simulation results of this study. So, taking the reduced
chemical potential $ |\Delta \mu|/(k_B T) \sim 0.19 \pm 0.03$ and
the cluster size $N \geq 3 N_c$, at which the steady-state growth
rate appears for the Dz-system, one obtains that the exponential
function in growth equation~(\ref{eq: growth_N}) becomes
approximately equal to unity, whereas $\upsilon_N^{(st)} \sim
g^+_{N_c}$ and, consequently, the thermodynamic impact to the
growth process is insignificant for the considered temperature
range $0 < T < T_g$.

Finally, the nucleation-growth rate characteristics and, in
particular, the kinetic coefficient $g_{N_c}^{+}$ are associated
with the mobility of particles and, thereby, with the diffusion
(as it is for $g_{N_c}^{+}$  in Kelton-Greer model, for example;
Eq.~(\ref{eq: Kelton_diffusion}) in Appendix) as well as with the
viscosity (see, for example, the principal equation of the kinetic
ballistic model, Eq.~(\ref{eq: ballistic}) in Appendix). In this
connection, the unified scenarios in temperature dependencies of
the growth and attachment rates presented in this paper provide
new important insight into the possibility to develop a general
model of viscosity suitable to all supercooled liquids regardless
of such the peculiarities as fragility, bonding and particles
interaction type~\cite{Nussinov}.

\section{Conclusions}

The main results of this study are the following:

(i) It is shown that the growth law of nucleus of near-critical
sizes can be represented as the cubic polynomial with the
coefficients dependent on the critical size, the attachment rate,
the chemical potential difference and the growth exponent -- i.e.
all the quantities incoming into original growth equation. The
growth law model is applied to evaluate crystal nuclei data
derived from molecular dynamics simulations for two different
crystallizing glassy systems at temperatures below $T_g$ via the
mean-first-passage-time analysis.

(ii) It is found that the crystal growth laws of the systems
follow the unified time-dependence, where the growth exponent is
constant  for each the system over whole the considered
temperature range.

(iii) The steady-state growth regime, where the growth rate is
independent of the cluster size, occurs when the size of growing
crystalline nucleus in a glassy system becomes two-three times
larger than a critical size.

(iv)  The results provide evidence that the rate characteristics
of the crystal growth process as  temperature dependent quantities
are reproducible within unified scaling relations. This finding is
supported for the steady-state growth rate and the attachment
rate. In particular, values of the reduced steady-state growth
rate taken from simulation results and experimental data for
different systems and plotted \textit{vs}. the scaled temperature
$\widetilde{T}$ are collapsed onto a single line for the
temperatures below $T_g$. In a similar matter, the reduced
attachments rates evaluated for the crystallizing glassy systems
and considered as functions of the reduced temperature
$\widetilde{T}$ follow the unified power-law dependence.

\section*{Appendix}

\subsection*{Zeldovich relation}

For a metastable system being at the low levels of supercooling
one can consider the dimensionless term
\begin{equation}
\frac{1}{k_B T} \frac{d\Delta G(N)}{dN}
\end{equation}
as a term taking small values less than unity. As a result,
Eq.~(\ref{eq: vel_2}) simplifies to the well known Zeldovich
relation for the growth rate:
\begin{equation}\label{eq: Zeldovich rel}
  \upsilon_N = - \frac{g^{+}(N)}{k_B T} \frac{d \Delta G(N)}{dN}.
\end{equation}
Zeldovich's relation~(\ref{eq: Zeldovich rel}) takes into account the fact that the growth rate is
resulted from both the thermodynamic and pure kinetic contributions, $d \Delta G(N)/dN$ and $g^{+}(N)$,
while their product defines completely
the temperature dependence as well as
the size dependence of the growth rate.
Nevertheless, the aforementioned condition
imposed on the thermodynamic term in Eq.~(\ref{eq: vel_2})
restricts application of the Zeldovich relation~(\ref{eq: Zeldovich rel})
to a system being at specific thermodynamic conditions.

\subsection*{Growth rate within the Wilson-Frenkel theory}

For the case, when
\begin{equation}
 \left ( \frac{N_c}{N} \right )^{1/3} \ll 1,
 \end{equation}
 that can be realized in the macroscopic limit, i.e. at the late stage of the growth of a solitary cluster or for growing crystalline slab,
 Eq.~(\ref{eq: vel_gen}) can be rewritten as
\begin{equation}\label{eq: WF_rel}
  \upsilon_N = g^{+}(N)\left \{ 1 - \exp \left [ - \frac{|\Delta \mu|}{k_B T}  \right ]  \right \}.
\end{equation}
For low and relatively moderate levels of supercooling, i.e. for
insignificant deviations from equilibrium, the next condition is
satisfied:
\begin{equation}\label{eq: cond_1}
  \frac{|\Delta \mu|}{k_B T} \ll 1.
\end{equation}
Then, Eq.~(\ref{eq: WF_rel}) takes the
form~\cite{Wilson_1900,Frenkel_1932}
\begin{equation} \label{eq: WF_rel_2}
  \upsilon_N \simeq  g^{+}(N) \frac{|\Delta \mu|}{k_B T}.
\end{equation}
Taking into account the thermodynamic identity
\begin{equation}\label{eq: equality}
  \frac{\Delta \mu}{k_B T} = \frac{l}{T_m}\frac{\Delta T}{k_B T} = \Delta S \frac{\Delta T}{k_B T},
\end{equation}
one comes to the known result, according to which the growth rate
$\upsilon_N$ within the Wilson-Frenkel
theory~\cite{Wilson_1900,Frenkel_1932,Kerrache} is proportional to
the supercooling $\Delta T = T_m -T$. Here, $l$ is the latent heat
of the transition and $\Delta S$ is the entropy change between the
crystalline and the liquid phases. This allows one to rewrite
Eq.~(\ref{eq: WF_rel_2})
\begin{equation} \label{eq: WF_rel_3}
    \upsilon_N = k^{WF} \Delta T,
\end{equation}
where $k^{WF}$ being the so-called interface kinetic coefficient.

As follows from the classical nucleation theory, since the chemical
potential difference $|\Delta \mu|$ is inversely proportional to the
critical size $N_c$ [see Eq.~(\ref{eq: critical size})], then
condition~(\ref{eq: cond_1}) corresponds to the large values of the
critical size, that indicates on the correspondence of Eq.~(\ref{eq:
WF_rel_2}) to the macroscopic growth rate.

\subsection*{Turnbull-Fisher model and its kinetic extension}

The Turnbull-Fisher model was specially adopted to describe the
crystal nucleation and growth in glasses~\cite{Turnbull_1949}. The
model was extensively studied by Kelton et al.~\cite{Kelton_1983}
and Greer et al.~\cite{Greer_1990}.

The structural transformations in glasses are driven rather by
kinetic than thermodynamic contribution. This means that the crystal
growth in a glass is largely defined by the diffusive processes. In
this model it is assumed that the coefficient $g^{+}(N)$ can be
taken in the form
\begin{equation}\label{eq: TF_kinetic_coef}
  g^+(N) \sim \exp\left [ - \frac{\Delta G(N+1) - \Delta G(N)}{2k_B T} \right ].
\end{equation}
Inserting Eq.~(\ref{eq: TF_kinetic_coef}) into Eq.~(\ref{eq: vel_3})
by analogy with Eq.~(\ref{eq: vel_2}) one obtains
\begin{eqnarray}
  \upsilon_N  &\sim& \exp\left [- \frac{1}{2 k_B T} \frac{d \Delta G(N)}{dN}  \right ] - \exp\left [ \frac{1}{2k_B T} \frac{d \Delta G(N)}{dN}  \right ] \\
       &\sim& 2 \sinh \left [ - \frac{1}{2k_B T} \frac{d \Delta G(N)}{dN}  \right ]. \nonumber
\end{eqnarray}\label{eq: TF_exp}
Then, taking into account Eq.~(\ref{eq: derivative of energy}) one
obtains the growth rate
\begin{equation}\label{eq: vel_TF}
  \upsilon_N \sim  2 \sinh \left \{  \frac{|\Delta \mu|}{2k_B T} \left [ 1 - \left( \frac{N_c}{N} \right)^{1/3}\right ]  \right \}  .
\end{equation}
As a result, expression for the Turnbull-Fisher
cluster-size-dependent growth rate can be written as
\begin{eqnarray} \label{eq: TF_complete}
  \upsilon_N &=& 2 g^+(N_c) \left ( \frac{N}{N_c} \right )^{2/3} \\
      & & \times \sinh \left \{  \frac{|\Delta \mu|}{2k_B T} \left [ 1 - \left( \frac{N_c}{N} \right)^{1/3}\right ]  \right \}, \nonumber
\end{eqnarray}

The Turnbull-Fisher model has been later extended by Kelton and
Greer. Namely, Kelton and Greer \cite{Kelton_JNCS_1986} expressed the
coefficient, $g^{+}(N_c) \equiv g^{+}_{N_c}$, in terms of the
ordinary diffusion $D$ as following
\begin{equation}\label{eq: Kelton_diffusion}
  g^{+}(N_c)= 24 \frac{D}{\lambda^2} N_c^{2/3},
\end{equation}
where $\lambda$ is the atomic jump distance. As a result,
Eq.~(\ref{eq: TF_complete}) takes the form
\begin{eqnarray}\label{eq: Kelton_complete}
  \upsilon_N &=& 48 \frac{D}{\lambda^2} N^{2/3}  \\
      & & \times \sinh \left \{  \frac{|\Delta \mu|}{2k_B T} \left [ 1 - \left( \frac{N_c}{N} \right)^{1/3}\right ]  \right \}. \nonumber
\end{eqnarray}

For the growth at macroscopic conditions, when the size of the
growing cluster is much larger than the critical size $N_c$,
Eqs.~(\ref{eq: TF_complete}) and (\ref{eq: Kelton_complete}) take
the forms,
\begin{equation}\label{eq: TF_macroscop}
  \upsilon_N = 2 g^+(N_c) \left ( \frac{N}{N_c} \right )^{2/3} \sinh \left \{  \frac{|\Delta \mu|}{2k_B T} \right \}
\end{equation}
and
\begin{equation}\label{eq: Kelton_macroscop}
  \upsilon_N = 48 \frac{D}{\lambda^2} N^{2/3} \sinh \left \{  \frac{|\Delta \mu|}{2k_B T} \right \}
\end{equation}
respectively.

It is easy to verify that at low levels of supercooling
Eqs.~(\ref{eq: TF_macroscop}) and (\ref{eq: Kelton_macroscop}) are
simplified to the next corresponding relations
\begin{equation}\label{eq: TF_near}
   \upsilon_N = g^+(N_c) \left ( \frac{N}{N_c} \right )^{2/3}\frac{|\Delta \mu|}{k_B T}
\end{equation}
and
\begin{equation}\label{eq: Kelton_near}
  \upsilon_N = 24 \frac{D}{\lambda^2} N^{2/3} \frac{|\Delta \mu|}{k_B T}.
\end{equation}
In Eqs.~(\ref{eq: TF_near}) and (\ref{eq: Kelton_near}) the
contribution responsible for to reproduce the thermodynamic aspect
of cluster growth remains the same, since the Kelton-Greer extension
concerns only the kinetic contribution of the Turnbull-Fisher model.

Further, taking into account equality (\ref{eq: equality}), one can see that
Eqs.~(\ref{eq: TF_near}) and (\ref{eq: Kelton_near})
yield the same linear dependence of the macroscopic growth rate
upon the supercooling, i.e. $\upsilon_N \propto \Delta T$,
\begin{eqnarray} \label{eq: Kelton_near_1}
  \upsilon_N &\propto& \frac{|\Delta \mu|}{k_B T} \\
      &\propto& \frac{l}{T_m}\frac{\Delta T}{k_B T} \nonumber \\
      &\propto& \Delta S \frac{\Delta T}{k_B T}, \nonumber
\end{eqnarray}
which is the same with Eq.~(\ref{eq: WF_rel_3}) derived within the Wilson-Frenkel theory.

\subsection*{Kinetic models}

There are some pure kinetic growth models, which are focused rather
upon the mechanisms of mass exchange between the cluster and the
mother phase~\cite{Shneidman_1992}. Hence, the $N^{2/3}$-dependence
for the coefficient $g^+(N)$ is suggested in the so-called ballistic
model~\cite{Volterra/Cooper_JNCS_1985}:
\begin{equation}\label{eq: ballistic}
  g^+(N) = b \frac{k_B T}{\eta \lambda^3} N^{2/3},
\end{equation}
where $\eta$ is the viscosity, and $b$ is a numerical constant.
Since for the three-dimensional growth one has $ dR/dt \propto
N^{-2/3} dN/dt$, then the ballistic model with Eq.~(\ref{eq:
ballistic}) correspond to the rate $g^{+}(N)$, which is actually
independent of the radius of the cluster, but depends on the
properties of the mother phase. As it was discussed in
Ref.~\cite{Volterra/Cooper_JNCS_1985}, such the scenario is
appropriate, in particular, for the case of the fluid droplet growth
in a supersaturated vapor.

Further, at the Ostwald ripening regimes of the growth kinetics
other situations appear~\cite{Langer_PRA_1980,Kukushkin_review,Rengarajan_2011},
where the rate $g^+(N)$ has the size-dependence of the form:
\begin{subequations}\label{eq: thetta}
  \begin{equation}\label{eq: exponent_two}
    g^+(N) \propto N^{1/3}
  \end{equation}
  and the rate $g^+(N)$ can be even independent of the size:
  \begin{equation}\label{eq: exponent_three}
    g^+(N) = \mathrm{const}.
  \end{equation}
\end{subequations}
As it is seen, relations (\ref{eq: ballistic}), (\ref{eq:
exponent_two}) and (\ref{eq: exponent_three}) can be generalized
into the power-law dependence~{\cite{Weinberg_2002,Shneidman_1992}}
\begin{equation}\label{eq: kin_gen}
  g^+(N) \propto N^{(3-p)/3},
\end{equation}
with $0 < p \leq 3$, where the integer values of the exponent, i.e. $p=1,\;2,\;3$, correspond to the models~(\ref{eq: ballistic}), (\ref{eq: exponent_two}) and (\ref{eq: exponent_three}), respectively.

\section*{Acknowledgments}
The authors thank E.~Zanotto, D.~Kashchiev, V. N.~Ryzhov, V. V.~Brazhkin for motivating discussions. The work is supported in part by the grant MD-5792.2016.2 (support for young scientists in RF) and by the project of state assignment of KFU in the sphere of scientific activities.

\end{document}